\pdfoutput=1
\documentclass[english,useAMS,usenatbib]{mn2e}
\usepackage[T1]{fontenc}
\usepackage{prettyref}
\usepackage{amsmath}
\usepackage{amssymb}
\usepackage{graphicx}
\makeatletter 
\providecommand{\tabularnewline}{\\}
\usepackage{epsfig}

\newcommand {\apgt} {\ {\raise-.5ex\hbox{$\buildrel>\over\sim$}}\ }
\newcommand {\aplt} {\ {\raise-.5ex\hbox{$\buildrel<\over\sim$}}\ } 

\usepackage{babel}

\usepackage{babel}
\makeatother
\usepackage{babel}
\begin{document}
\title[Early phases of AGN jets]{3D simulations of the early stages of AGN jets: geometry, thermodynamics and backflow}

\author[S. Cielo et al.]{S.~Cielo$^{1}$%
\thanks{e-mail: cielo@mpia.de%
}, V.~Antonuccio-Delogu$^{2,\,3,\,4}$, A.V. Macci\`o$^{1}$, A.D. Romeo$^{5}$
and J. Silk $^{6,\,7,\,8}$ \\
$^{1}$ Max-Planck Institute for Astronomy, K\"onigstuhl 17, 69117 Heidelberg,
Germany\\
$^{2}$ INAF, Catania Astrophysical Observatory, Via S. Sofia 78,
I-95126 Catania, Italy\\
$^{3}$ Institute for Theoretical Astrophysics, University of Heidelberg,Albert-Ueberle-Str. 2 and Philosophenweg 12, 69120 Heidelberg, Germany\\
$^{4}$ Scuola Superiore di Catania, Via Valdisavoia 9, 95123 Catania,
Italy\\
$^{5}$ Departamento de Ciencias Fisicas, Universidad Andres Bello,
Av. Republica 220, Santiago, Chile\\
$^{6}$ 
Institut d'Astrophysique de Paris - 98 bis boulevard Arago -
75014 Paris, France\\
$^7$The Johns Hopkins University, Department of Physics and Astronomy,
3400 N. Charles Street, Baltimore, MD 21218, USA\\ \indent
$^8$Beecroft Institute of Particle Astrophysics and Cosmology, Department of Physics,
University of Oxford, Denys Wilkinson Building,\\ \indent 1 Keble Road, Oxford OX1 3RH, UK
}

\date{Accepted 2014 January 17. Received 2014 January 16; in original form 2013 November 19}
\maketitle
\pagerange{\pageref{firstpage}--\pageref{lastpage}} \pubyear{2013}
\label{firstpage} 
\begin{abstract}
We investigate the interplay between jets from Active Galactic Nuclei (AGNs) and the surrounding InterStellar Medium (ISM) through full 3D, high resolution, Adaptive Mesh Refinement simulations performed with the {\sc flash} code. 
We follow the jet-ISM system for several Myr in its transition from an early, compact source to an extended one including a large cocoon.
During the jet evolution, we identify three major evolutionary stages and we find that, contrary to the prediction of popular theoretical models, none of the simulations shows a self-similar behavior.
We also follow the evolution of the energy budget, and find that the fraction of input power deposited into the ISM (the \emph{AGN coupling constant}) is of order of a few percent during the first few Myr. This is in broad agreement with galaxy formation models employing AGN feedback. However, we find that in these early stages, this energy is deposited only in a small fraction ($< 1\%$) of the total ISM volume.
Finally we demonstrate the relevance of \emph{backflows} arising within the extended cocoon generated by a relativistic AGN jet within the ISM of its host galaxy, previously proposed as a mechanism for self-regulating the gas accretion onto the central object. These backflows tend later to be destabilized by the 3D dynamics, rather than by hydrodynamic (Kelvin-Helmholtz) instabilities. Yet, in the first few hundred thousand years, backflows may create a central accretion region of significant extent, and convey there as much as a few millions of solar masses.
\end{abstract}
\section{Introduction} \label{sec:intro}
 \par Active Galactic Nuclei (AGNs) are responsible for highly energetic outflow events, powered eventually by matter inflow into the gravitational potential of supermassive black holes (SMBHs), believed to be present at the centers of many, if not all, galaxies. 
Feedback from active galactic nuclei plays an important role in the energy balance of their host galaxies: AGNs are often invoked in theoretical models as a heating source capable of quenching the star formation (SF) in high-mass galaxies (\emph{Negative feedback}: \citealp{2006MNRAS.365...11C}, \citealp{2008MNRAS.391..481S}, \citealp{2009MNRAS.396...61T}, \citealp{2012ARA&A..50..455F}). Nevertheless, AGNs are sometimes reported to have the opposite effect (\emph{Positive feedback}, see e.g. \citealp{Gaibler:2011hc}).

According to our current understanding, the scenario is quite complex in part because several different outflow mechanisms have been observed/proposed to possibly originate from an AGN, mostly determined by the rate of the mass inflow onto the SMBH that makes the galactic nucleus active. It is possible to distinguish (see e.g \citealp{2012MNRAS.419.2797F}): a \emph{quasar} (radiative) mode powered by a high accretion rate; a \emph{jet} (or \emph{kinetic} or \emph{radio}) mode, when having a low one.
Among the various outflow regimes, we focus on the ``jet mode'', in which two very energetic jets of relativistic matter are shot from the nucleus in opposite direction. Such jets undergo strong interaction/mixing with the surrounding gas, immediately becoming mass dominated (at least, for the scales we resolve)." They can propagate up to several tens or hundreds of kpc, carving a \emph{cocoon} in the surrounding gas, and eventually generating the bipolar-shaped, very luminous emission associated with  \emph\emph{radio galaxies}, such as  \emph{Faranoff-Riley II} (FRII) galaxies. Besides, the jet mode might be the only mechanism active for long enough to have significant impact on galactic SF. 

The study of the interactions of an active central object with the interstellar medium (ISM) or, on larger scales, the circumgalactic medium (CGM), is thus an interesting and promising investigation topic.
Since the paper by \cite{1998A&A...331L...1S}, the importance of AGN Jet feedback received strong support from both theoretical (e.g. \citealp{Sutherland:2007rz,Gaibler:2011hc}) and observational (e.g. \citealp{2006Natur.442..888S,2009A&A...507.1359E}) perspectives. Nevertheless, substantial limitations in both fields give rise to many highly debated -yet unanswered- questions. Among the most important of these questions, some concern, generally speaking: 
  \begin{enumerate}
   \item the first evolutionary stages of these objects: if and how observed compact radio sources\footnote{such as CSSs, \emph{Compact Steep Spectrum sources} and GPSs, \emph{Gigahertz Peaked Spectrum sources}} eventually evolve into extended sources such as FRII galaxies;
    \item whether the coupling between the AGN and the rest of its host galaxy or halo is high enough to allow for substantial energy transfer, and how this is achieved;
    \item the jet physical composition and thermodynamic state after its contacts with the ISM, which also greatly affects the previous point;
    \item whether some self-regulation mechanism is driving the alternation between active and passive phases, e.g. by regulating the central mass inflow rate.
  \end{enumerate}

\begin{figure}
  \centering
  \includegraphics[width=1\columnwidth]{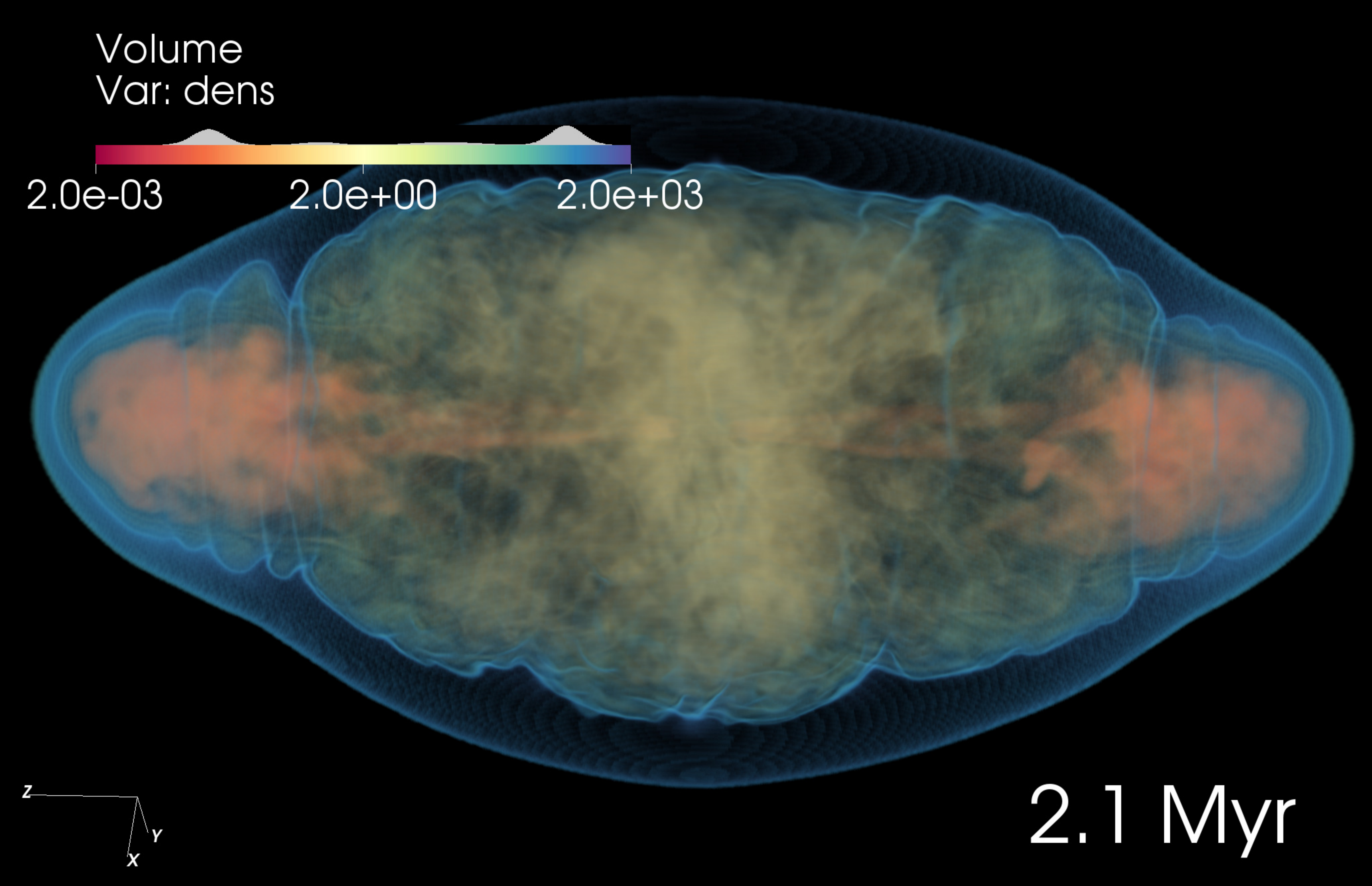}
  \caption{3D ray-tracing density rendering of run dj-250p6, at $2.1$ Myr. The legend shows, above the color code, the corresponding opacity used for the ray tracing. This combination was chosen to highlight different regions: bow shock (blue), cocoon material (yellow), jets and lobes (orange). The external ISM is not shown for simplicity.}
  
  \label{fig:3Drender}
\end{figure}

For decades, several steps have been taken in theoretically models or implementations of jet feedback in coherently simulated scenarios to investigate these problems. Most difficulties arise from AGNs being intrinsically multi-scale objects, in which one has to model and resolve several physical processes, such as hydrodynamics, radiative cooling, gravity and star formation.

As for the early evolution of radio galaxies, several analytic models have been proposed. The model by \cite{1991MNRAS.250..581F} and later extended by \cite{1997MNRAS.286..215K,2002MNRAS.335..610A}\, describes the \emph{global average} properties of cocoon dynamics in term of this expansion, predicting that the cocoon's expansion is self-similar during most of its life. However, this self-similar expansion model leaves out questions related to the internal dynamics and thermodynamics of the jet-cocoon system. Under realistic circumstances, self-similarity may not hold, and indeed it is not likely to. \cite{2005MNRAS.364..659K} and \cite{2008ApJ...687..141K} propose a model for expansion in a non-uniform ISM which accounts separately for the cocoon transverse expansion, thus  not implying (albeit not excluding) self-similarity, and test it against young radio galaxies.

Important results have been found also through simulations: for instance \cite{Sutherland:2007rz} described precise evolutionary stages for the jet/cocoon system; on larger scales \cite{2006MNRAS.373L..65H} reproduced X-ray luminosities of observed bright sources such as Cyg$\alpha$.

Other authors such as \cite{2012ApJ...757...65S,2013ApJ...774...12F} provided insights on jet-launching mechanisms from magnetized accretions disks;  \cite{2012MNRAS.423.3083M} studied accretion flows on spinning black holes in general relativistic magnetohydrodynamic (GRMHD) simulations, finding highly magnetized states that drive inflows and generate stable relativistic jets in agreement with the \emph{Blandford-Znajek} (BZ) jet model (\citealp{Blandford:1977ds}). Also \cite{2013arXiv1307.1143S} and \cite{2013MNRAS.436.3741P} used GRMHD simulations to investigate different disk and magnetic field configurations, that again appear consistent with the BZ model.

The energy balance of the jet feedback, and its effects on star formation have been studied through hydrodynamic simulations: \cite{Gaibler:2010ij} found asymmetries between the two lobes to be significant; \cite{Gaibler:2007fv} and \cite{Gaibler:2011hc} studied the interaction with the ISM, with particular attention to star formation, favouring a positive feedback scenario.
\cite{2009MNRAS.396...61T} obtained color indexes from 2D simulations, which indeed suggested a positive feedback initial transient phase, triggered by mechanic gas compression, but overall negative feedback (due to cold gas heating and clump destruction) after a few tens of Myr. Their predicted colours agree well with observed spheroidal galaxies that had AGN-related events.

The importance of self-regulation in AGN environments has been recently highlighted by \cite{2011MNRAS.411..349G,2011MNRAS.415.1549G}, who studied in detail the interplay of AGNs with cold/hot accretion. \cite{2010MNRAS.405.1303A} have demonstrated the rise of \emph{backflows} within the global circulation inside a \emph{cocoon} (see \citealp{2007MNRAS.382..526P,2008A&A...488..795R,2010ApJ...709L..83M}) generated by the interaction of an AGN jet with the ISM of its host galaxy. Such backflows have been noticed since the first numerical simulations \citep{1982A&A...113..285N, 2010MNRAS.405.1303A}; they do act as a self-regulation mechanism, driving gas back towards the meridional plane in a two-lobe system. In the aforementioned work by \protect\cite{2010MNRAS.405.1303A}, these backflow are a consequence of local discontinuities in the entropy. In that framework, \emph{Crocco's theorem} states that such discontinuities act as a vorticity source term. Since very large entropy discontinuities are present at the interface between the jet and the Hot Spot (HS), strong vorticity may be naturally generated in FRII sources through this mechanism, driving the backflows. This latter paper also observed how the backflows were stable for most of the evolution of the jet-cocoon system, but the scope of this work was restricted by the fact that the simulations were only in 2D.

One problem with analytic models is that it is difficult to predict which among the jet/cocoon internal dynamics are relevant, and properly include them. Though for example \cite{1997MNRAS.286..215K} include jet recollimation shocks, other dynamics may be in play: jet propagation generates turbulence within the cocoon and, if this turbulence is isotropic, an isotropic \emph{turbulent pressure} $p_{t}$ arises, which adds to the gaseous thermal pressure\footnote{Throughout this paper we will use uppercase $P$ to denote power, and lowercase $p$ to denote pressure} $p_{g}$. Also, the results depend on the assumption for the gas distribution in the ISM. This is true also for simulations: changing for instance the distribution of the cold (or warm) gas phase affects the results heavily; also, there is so far no general agreement on feedback outcomes or indications for a unitary picture of AGN jets.

In the present paper we focus our attention on the \emph{internal} properties of the jet-cocoon system in a new set of full 3D, \emph{Adaptive Mesh Refinement} (AMR) simulations, where we provide detailed cocoon shape measurements We try to compensate our ignorance on the jets' physical composition by running different ``families'' of simulations varying the jet/ISM density contrast and relating this to the cocoon shape.  We then investigate the jet/cocoon thermodynamical properties: how pressure shares between turbulent and isothermal pressures, whether turbulent pressure affects the global dynamics of the cocoon, and how this is linked to the evolution of the system. We analyse the cocoon/ISM energy balance, including the energy deposition in the form of mechanical ``$p\,dV$'' work and likewise ``$T\,dS$'' exchanged heat.

In the end,  we present updated results on backflows: we study how much they can contribute to supply the accretion region around the SMBH with gas and energy. The backflow carries very hot, high pressure gas; thus, it can heavily affect the \emph{circumnuclear star formation} and the properties of the accretion disc. 

The simulation setup we use is introduced in Section \ref{sec:setup}, while an overview of the typical run evolution is in Section \ref{sec:Evolutionary-stages}. Section \ref{sec:geometry} and \ref{sec:thermodynamics} are dedicated to cocoon geometry and thermodynamics, respectively. We deal with backflows in Section \ref{sec:backflow}, while in Section \ref{sub:backflowStability} we also investigate their \emph{stability}, which affects this feedback mechanism. Section \ref{sec:discussion} contains the discussion.

\section{Simulation setup and runs} \label{sec:setup}
\subsection{Simulation volume}\label{sub:simVolume}
\begin{table*}
  \begin{tabular}{cccccccc}
  \hline 
  name & $\sigma_{V}\,/\, \rm{km\, s^{-1}}$ & $\rho_{halo}^{c}\,/\, \rm{MFLY\: Mpc^{-3}}$ & $\rho_{jet}\,/\, \rm{MFLY\: Mpc^{-3}}$ & $p_{jet}\,/\, \rm{PFLY}$ & $\log(\frac{P_{jet}}{\rm{W}})$ & $\mathcal{M}_{int}$ & $t_{MAX}\ /\ \rm{Myr}$\tabularnewline
  \hline 
  150p3 & 150 & $2311.3$ & $23.113$ & $10^{3}$ & $37.6727$ & 164.4 & 2.2\tabularnewline
  150p5 & 150 & $2311.3$ & $23.113$ & $10^{5}$ & $37.6727$ & 16.4 & 3.1\tabularnewline
  150p6 & 150 & $2311.3$ & $23.113$ & $10^{6}$ & $37.6727$ & 5.2 & 3.6 \tabularnewline
  200p5 & 200 & $1748.28$ & $17.4828$ & $10^{5}$ & $37.9359$ & 19.2 & 2.3\tabularnewline
  200p6 & 200 & $1748.28$ & $17.4828$ & $10^{6}$ & $37.9359$ & 6.1 & 2.7\tabularnewline
  250p5 & 250 & $1410.2$ & $14.102$ & $10^{5}$ & $37.14$ & 21.7 & 1.9\tabularnewline
  250p6 & 250 & $1410.2$ & $14.102$ & $10^{6}$ & $37.14$ & 6.9 & 3.8\tabularnewline
  \hline 
  d+200p5 & 200 & $17482.8$ & $17.4828$ & $10^{5}$ & $37.9359$ & 19.2 & 1.4\tabularnewline
  d+200p6 & 200 & $17482.8$ & $17.4828$ & $10^{6}$ & $37.9359$ & 6.1 & 2.8\tabularnewline
  d+250p5 & 250 & $14102$ & $14.102$ & $10^{5}$ & $38.14$ & 21.7 & 5.7\tabularnewline
  d+250p6 & 250 & $14102$ & $14.102$ & $10^{6}$ & $38.14$ & 6.9 & 5.5\tabularnewline
  \hline 
  dj-200p5 & 200 & $1748.28$ & $1.74828$ & $10^{5}$ & $37.9359$ & 13.1 & 2.1\tabularnewline
  dj-200p6 & 200 & $1748.28$ & $1.74828$ & $10^{6}$ & $37.9359$ & 4.13 & 2.5\tabularnewline
  dj-250p6 & 250 & $1410.2$ & $1.4102$ & $10^{6}$ & $38.14$ & 4.67 & 3.1\tabularnewline
  \hline 
  \end{tabular}
  \caption{Defining parameters of our simulation runs: name of the run, halo central velocity dispersion $\sigma_{V}$, central (Dark Matter) halo density $\rho_{halo}^{c}$, jet density $\rho_{jet}$, jet pressure $p_{jet}$ (a proxy for its internal energy $e_{jet}$), jet injection mechanical power $P_{jet}$, the related jet's internal Mach number $\mathcal{M}_{int}$, final simulation epoch $t_{MAX}$. The other halo and jet physical parameters are all uniquely determined from these ones by using the scaling relations discussed in Section \ref{sec:setup}. The first seven runs make up our \emph{fiducial} family, and have their density contrast $\rho_{halo}^{c}/\rho_{jet}$ set to $100$. The prefix ``d+'' denotes the \emph{denser ISM} runs, that have a ten times larger $\rho_{halo}^{c}$. Prefix ``dj-'' denotes the \emph{light jet} family, that  has instead the $\rho_{jet}$ reduced by the same factor. Thus in both cases $\rho_{halo}^{c}\,/\,\rho_{jet}=1000$.  We remind that MFLY\(\simeq5.23\times10^{12}\rm{M_{\odot }}\) and PFLY$\simeq1.80\times10^{-15}$Pa.}
  \label{tab:run}
\end{table*}
The initial setup is devised to model the environment of a spheroid, which could either be an early-type galaxy or a pseudo-bulge component of a late-type one, with an isothermal gaseous profile embedded in equilibrium within a Dark Matter halo. In the present work we are mainly interested in modelling the large-scale properties of the jet-cocoon system, thus we do not put a disc of cold clouds as, for instance, in \citet{2012ApJ...757..136W}. A disc indeed is not likely to affect large-scale properties, as found for instance in \cite{Gaibler:2011hc}. 

Our jets propagate into a hot, isothermal ($T=10^{7}$K), low-density ISM, representative of the diffuse ISM of the spheroid described above. The spheroid is not rotating, in order to test the scenario described in \ref{sec:intro} in the most straightforward way. We wrote our setup in 	extsc{FLASH}, a block-structured, adaptive mesh-refinement hydrodynamic code (see \citealp{Fryxell:2000fk}). We adopt a rectangular simulation box, with a volume of $[60\times60\times(2\times60)]\, \rm{kpc^{3}}$, so that the jet can be shot from the centre and propagate parallel to the longest side. We had 	extsc{FLASH} dealing with it as the juxtaposition of two cubic cells, through the use of the \emph{Multigrid/Pfft} hydro-solver, capable of dealing with simulation boxes of such composite (non-cubic) shape. 
\emph{Multigrid} refers to the algorithm capability of dealing with grids with non-uniform resolution (as many in 	extsc{FLASH} can); \emph{Pfft} explicates that Fourier transforms are executed wih parallel solvers on the whole domain, instead of serial solvers applied block-by-block by local processors. This improves the algorithm scalability and fixes an important load imbalance of some original Multigrid methods. We use the 	extsc{FLASH} default \textit{outflow} boundary conditions on all the sides of the box. 

We take advantage of the 	extsc{FLASH} Adaptive Mesh Refinement (AMR) capabilities to achieve high spatial resolution: in the 	extsc{FLASH} AMR implementation, the simulation volume is recursively divided on-the-fly in blocks, splitting in half along each direction at each level of refinement (e.g. in 3D every block is split in 8 equal parts). This goes on until the user-set refinement criteria are no longer verified (i.e. gradients calculated on the grid are not too large), or the chosen maximum refinement level $l_{max}$ is reached.
We use the 	extsc{FLASH}'s default refinement criteria, based on Loehner's error estimator, set to $0.8$ for refinement and $0.6$ for de-refinement.  In all the runs we show in this work (Table~\ref{tab:run}) we put $l_{max}=9$. This implies that the smallest block will have a volume $L_{b}^{3}/(2^{9})^{3}=7.45\times10^{-9}L_{b}^{3}$.
Each block is further divided in cells: we use $8^{3}$ cells per block. "In this way, we have a smallest cell size of $6\times10^{4}pc/8/2^{9}\simeq14.6$ pc, sufficient to resolve small scale turbulence creation/dissipation. 

We adopt, as the  internal unit system, the \textit{FLY system} ($L_{0}=1\, \rm{Mpc} \,t_{0}=2/3H_{0},M_{0}=5.229\times10^{12}\rm{M_{\odot}}$, so that:\,$GM_{0}t_{0}^{2}/L_{0}^{3}=1$, see \citealp{2003CoPhC.155..159A}) in order to avoid numerical truncation problems which may arise in SI or CGS units.

Our physical setup includes gravity from an external, static dark matter halo having a NFW \citep{1996ApJ...462..563N} density profile, plus the contribution of a central SMBH. As for the hydrodynamic component, we model a single-fluid multi-phase gas. A hot ISM phase is specified as an initially isothermal ($10^{7}$ K) plasma, embedded in gravitational equilibrium within the NFW external potential, and subject to radiative cooling. The other component we adopt are the jets: from the very centre of the halo we launch two jets in opposite directions, modeled as a uniform, cylindric constant source term of about $30$ pc diameter (a few cells). 

We also include plasma energy loss by radiative cooling, implemented as prescribed by \cite{SutherlandDopitaCooling}, whose tables have been extended to higher plasma temperature, up to $10^{12}$K \citep[][Appendix B]{2008MNRAS.389.1750A}.

We then follow the evolution of the jets+ISM system for several Myr, in order to observe the early stages of their life, and the transition phase to larger sources such as the Medium-sized Symmetric Objects (MSOs) or fully-developed FRII galaxies. A visual impression is given in the 3D rendering in Figure~\ref{fig:3Drender}.

\subsection{Scaling relations, host galaxy and jet parameters}\label{sub:scaling}
Our setup can be seen as the 3D extension of the one adopted in \cite{2008MNRAS.389.1750A,2010MNRAS.405.1303A}.
For our first family of runs (which we call the \emph{fiducial} runs in Table~\ref{tab:run}), the physical parameters of the halo and jet are all chosen as in \cite{2010MNRAS.405.1303A}, taking just the average value of the scaling relations cited therein, with the aim to describe low/medium power FRII radiogalaxies. Everything is once again calibrated on the halo central velocity dispersion $\sigma_{V}$.

The virial mass $M_{vir}$ of the host halo is assumed to scale with $\sigma_{v}$\,as in Figure 3 of \cite{2006ApJ...648..826L}
\begin{equation}
  M_{vir}=2.57\times10^{12}\,\left(\frac{\sigma_{V}}{200\, \rm{km\, s^{-1}}}\right)^{2.99\pm0.15}\rm{M_{\odot}}
\end{equation}
which we use in turn to predict the halo concentration parameter $c_{NFW}$ as in \cite{Maccio21122008}:
\begin{equation}
  c_{NFW}=9\left(\frac{M_{vir}}{M_{*}}\right)^{-0.13},\ M_{*}=1.5\times10^{13}h^{-1}\rm{M_{\odot}}
\end{equation}
From these two parameters, one can analytically calculate the halo's virial radius $r_{vir}$, taken as the radius for which
\begin{equation}
  M_{vir}=200\times\frac{4}{3}\pi{r_{vir}}^3\rho_{crit}
\end{equation}
where $\rho_{crit}$ is the critical density of the Universe. We do not aim to be cosmologically accurate, so we just use a $\Lambda$-CDM cosmology with reduced Hubble constant $h=0.7$. We finally calculate the halo central (dark matter) density from the definition of the $c_NFW$ parameter: 
\begin{equation}
	\rho_{halo}^{c}=M_{vir}/\left(4.0\,\pi\,{f_c}\,r_S^3\right)
\end{equation}
being $r_S=r_{vir}/c_{NFW}$ the \emph{scale radius} of the halo and 
\begin{equation*}
	f_c=\log\left(1+c_{NFW}-\frac{c_{NFW}}{1 +c_{NFW}}\right).
\end{equation*}
For the central black hole, we assume that its mass $M_{BH}$ scales with $\sigma_{V}$ according to the relation found by \cite{ferrarese2000apjl}:
\begin{equation}
  M_{BH}=\left(1.2\pm0.2\right)\times10^{8}\left(\frac{\sigma_{V}}{200\, \rm{km\, s^{-1}}}\right)^{3.57\pm0.3}\rm{M_{\odot}}
\end{equation}
and finally, for the jet total mechanical power $P_{jet}$ we follow Eq. 9 of \cite{liu2006jet} (where we put $\lambda=L_{bol}/L_{edd}=0.1$):
\begin{equation}
  log_{10}(P_{jet})=-0.22+0.59\, log_{10}\left(\frac{M_{BH}}{M_{\odot}}\right)+33.48
\end{equation}
where $P_{jet}$ is expressed in Watt; this assumes that the jet power ultimately comes from BZ process, as supported by recent GRMHD simulations (e.g. \citealp{2012JPhCS.372a2040T}).\\\indent
We have then chosen three cases, namely the ones corresponding to $\sigma_{V}=150,\,200\,\mbox{and}\,250\, \rm{km\, s^{-1}}$. 
\begin{figure*}
    \centering
    \includegraphics[width=0.249\textwidth]{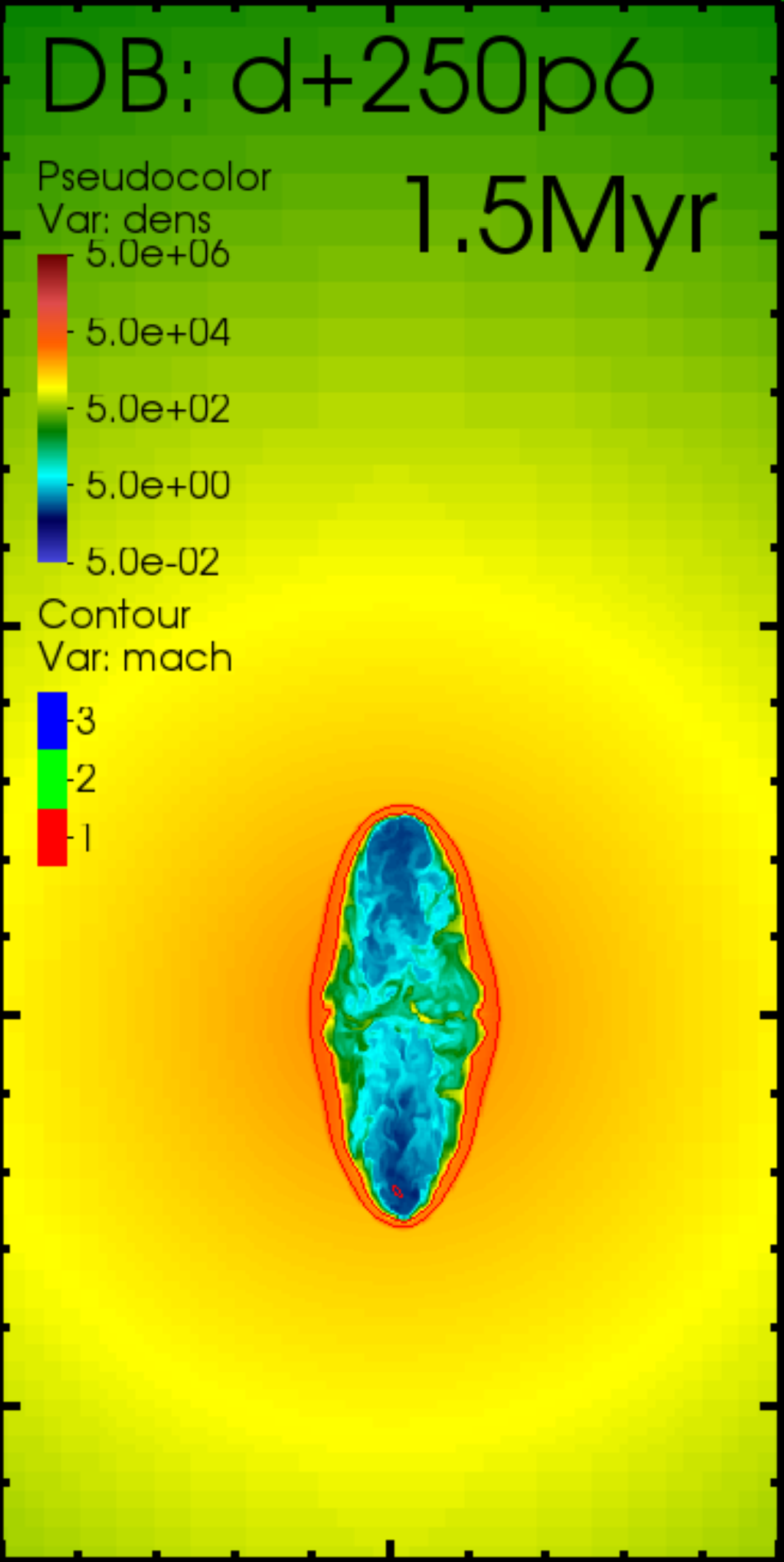}\includegraphics[width=0.249\textwidth]{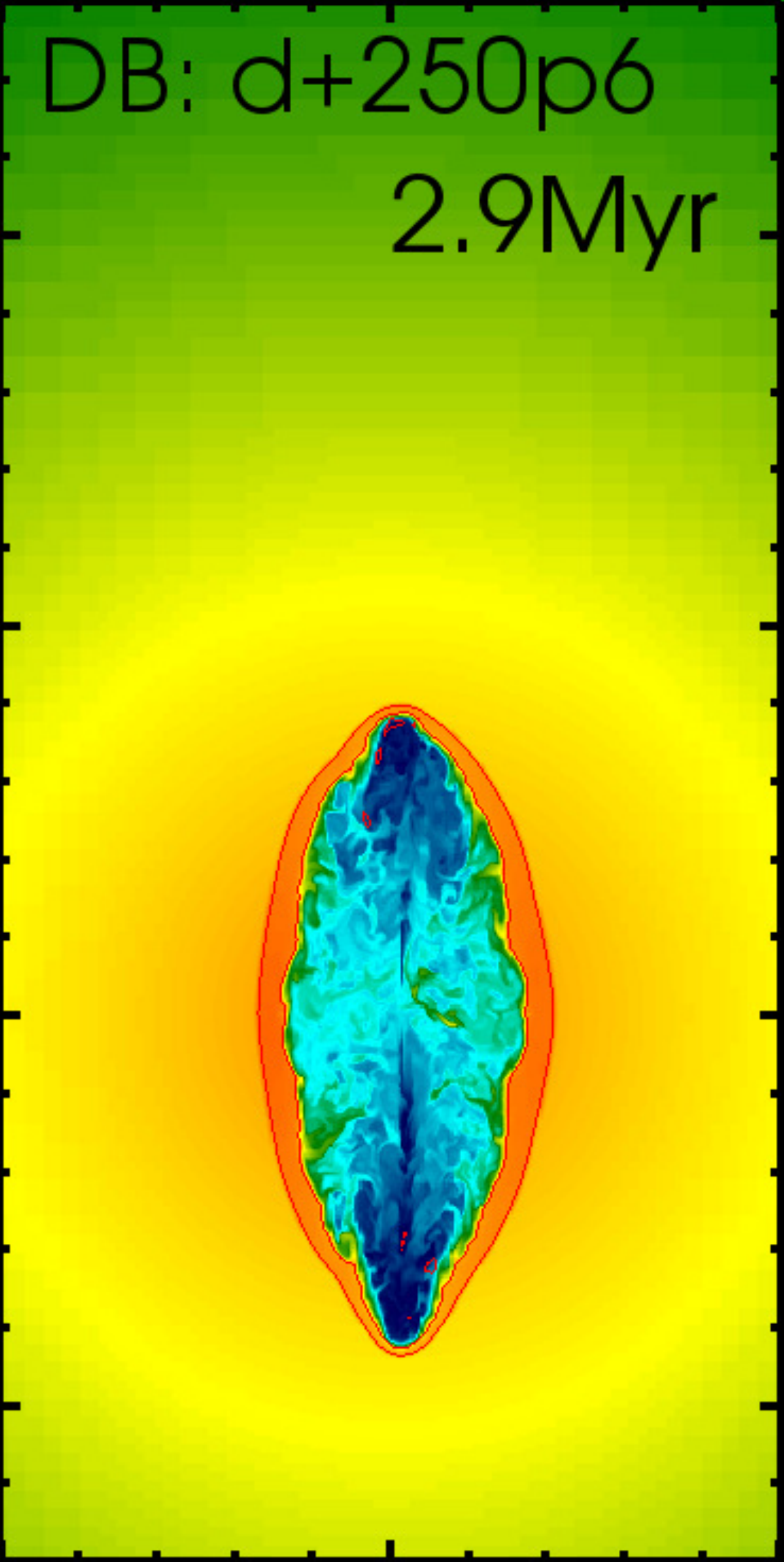}\includegraphics[width=0.249\textwidth]{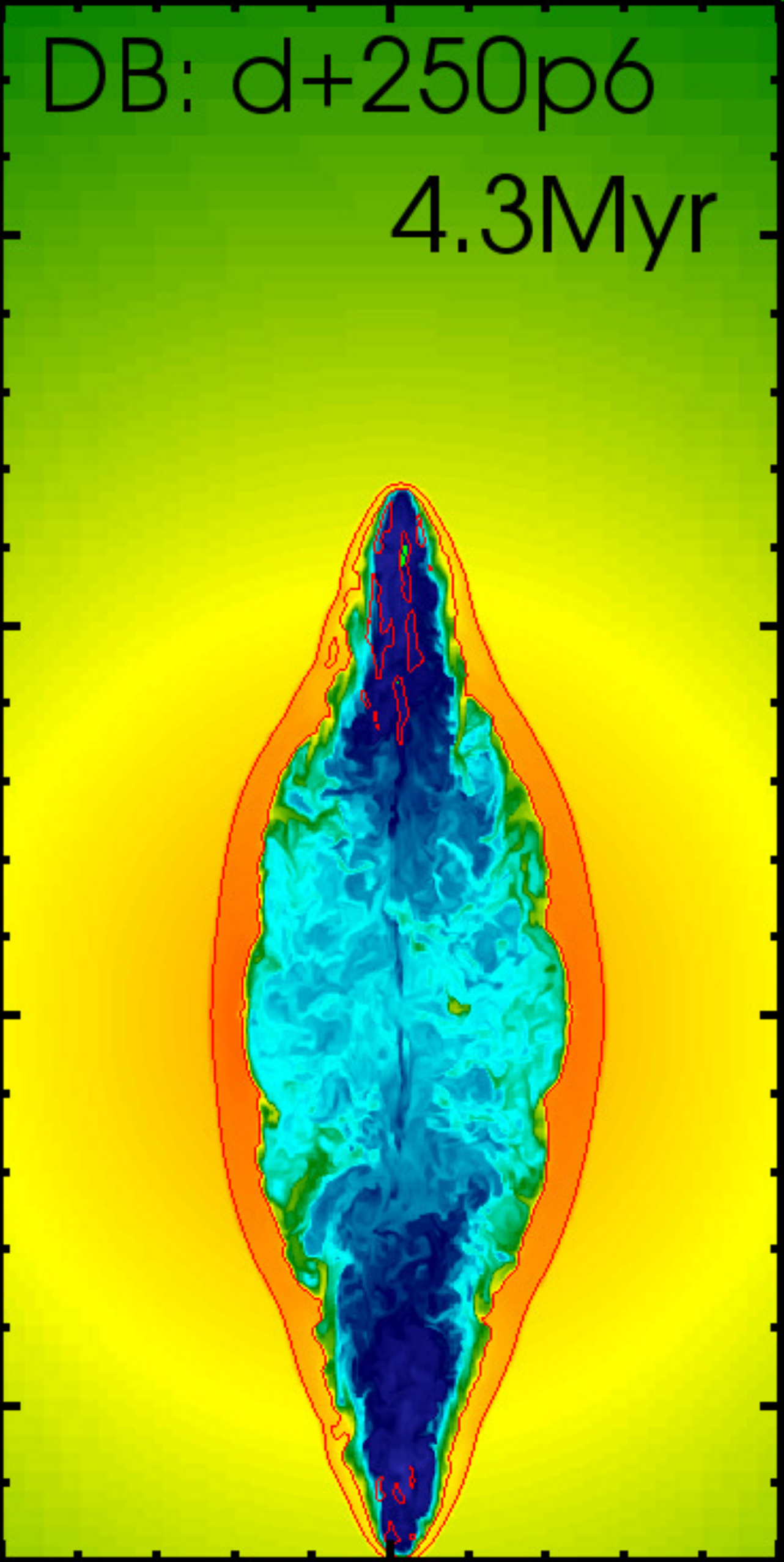}\includegraphics[width=0.249\textwidth]{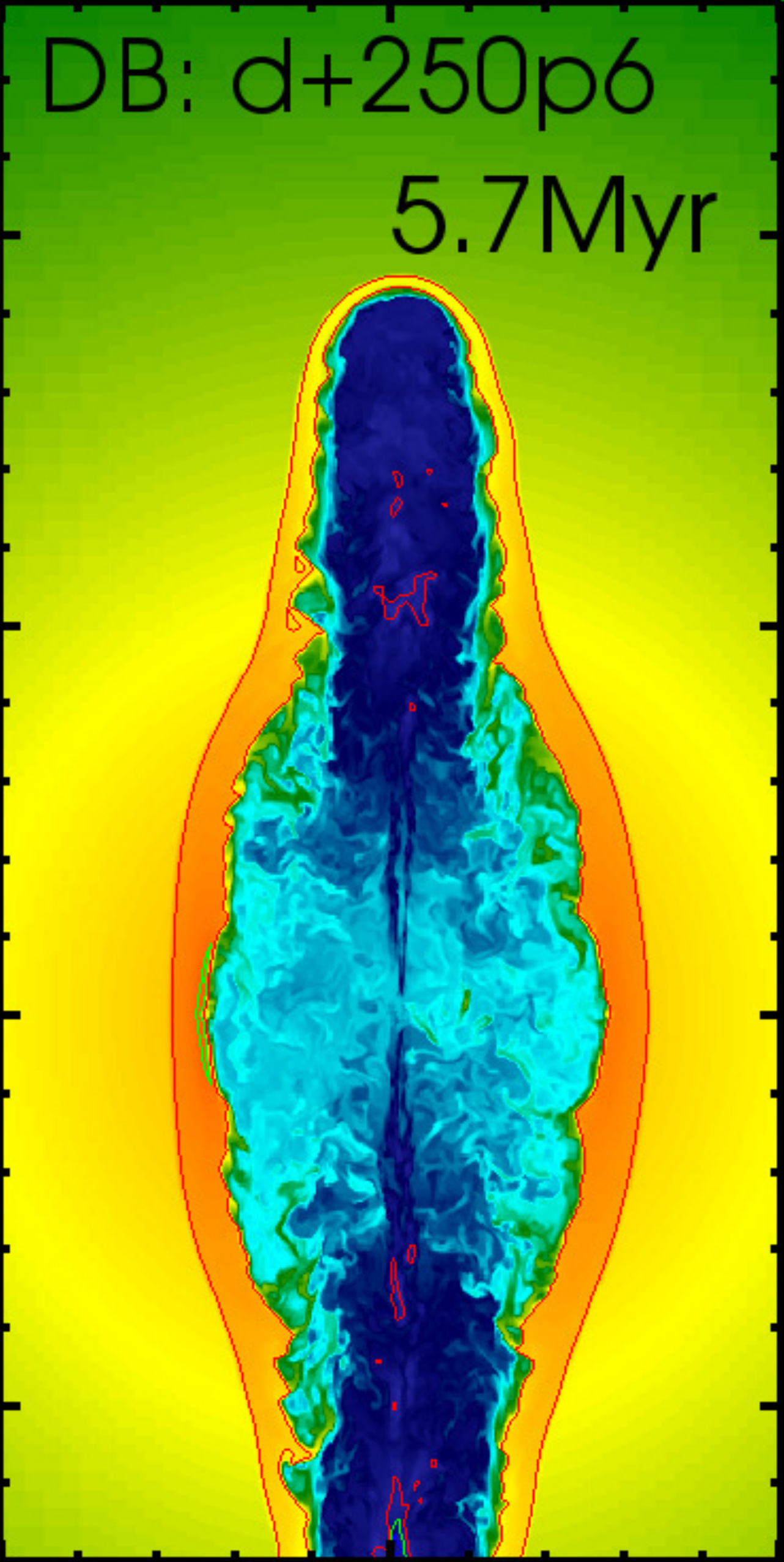}
    \caption{Central plane density slices  of run d+250p6, showing four evolutionary stages, from $1.5$ to $5.7$ Myr. The contours show the fluid Mach number. Size: $10\times20$ kpc (small ticks every kpc, large ones every $5$ kpc). The halo centre is located at $7$ kpc from the bottom edge. This run has been chosen for having the highest simulation age, and for showing clearly many important features, such as cocoon formation and penetration.}
    \label{fig:evolution}
\end{figure*}
In the fiducial runs, the jet density $\rho_{jet}$ is assumed to be $1/100$ of the central halo density; the source velocity $v_{jet}$ is thus determined by $P_{jet}$. In order to completely specify the source thermodynamic state, we have to also set the pressure $p_{jet}$ of the jet plasma (or, equivalently, its temperature $T_{jet}$). This variation corresponds to different internal Mach numbers $\mathcal{M}_{int}$ of the jet, so by varying this parameter we can explore different regimes. 

We limit ourselves to the supersonic case -where the system is not supposed to be very sensitive to this parameter- pushing towards the edge of the transonic case ($\mathcal{M}_{int}\sim4$). In order to approach these values, we had to set $p_{jet}=10^{6}$PFLY (i.e .$1.8\times10^{-9}$Pa). For the highly supersonic runs, we have safely set a value of $p_{jet}$ ten times smaller.

We then designed a \emph{light jet} family of runs, denoted by the prefix ``dj-'' and obtained by decreasing the jet density in order to have a density contrast of  $1/1000$. In turn we raised $v_{jet}$ to still match the same $P_{jet}$. Jets of such densities are considered \emph{very light}, yet they are commonly adopted in order to have less elongated cocoons \citep{Gaibler:2007fv}. Thus, we decided to run a third family, the \emph{denser ISM} family (``d+'' prefix), where the same density contrast is obtained by raising the ISM central density by a factor of ten. 
\section{Evolutionary stages} \label{sec:Evolutionary-stages}
The runs listed in Table \ref{tab:run} show different evolution paths, yet we can identify three main evolutionary phases.
In Figure \ref{fig:evolution} we show one significant run for illustration purposes (run d+250p6), while differences among runs are presented in  Figure \ref{fig:geoCocoon}. 

Furthermore, some runs more than others show asymmetry between the two jets/cocoons, so that the two halves of the cocoon can be at different stages at the same time. But these asymmetries are never very significant, as they get less pronounced with increasing simulation time $t_{age}$. This is simply a consequence of the development of turbulence, as asymmetry occurs apparently ``at random'' for what concerns direction, timing and intensity.

We can identify three main phases in the evolution of the jet-cocoon system:
\begin{enumerate}
  \item \textbf{``C'' phase: Cocoon and hotspot formation }- At the very beginning of the simulation ($t\lesssim0.1$ Myr) the (highly supersonic) jet  produces -through a strong shock- a \textit{hotspot} (HS) right where the dense ISM is frontally hit. The HS spans less than $1$ kpc in size, and it is characterized by very high temperature ($T\gtrsim10^{11}$ K) and pressure. \\Meanwhile, an ellipsoidal \textit{bow shock} region starts to expand from the centre, moving at approximately its local speed of sound. This wavefront wipes and accumulates gas in a thin (roughly a few hundreds of pc) layer, that ``shields'' the ISM from the hot jets. This bow-shock fronts continues to propagate in all directions, isolating an ``inner'' region: the cocoon. In the following we will (as in e.g. \citealp{1991MNRAS.250..581F}, \citealp{2005MNRAS.364..659K}) treat the cocoon as a two-axial ellipsoid. We also follow separately its two halves, accounting for asymmetry between the two jets. We will indicate the semi-major axis with $r_{HS}$, as it coincides with the distance of the HS from the centre\footnote{Actually, having a bipolar jet, we take the mean of these two distances for each run; see Section \ref{sec:geometry} for a formal definition of these parameters}. The semi-minor axis will be $r_C$, for \emph{cocoon radius}.\\
  This phase lasts about $0.3$ Myr for the fiducial runs, but it can last up to $\leq1$ Myr if the density contrast is enhanced. In Figure \ref{fig:evolution}, a late C phase is shown in the first panel. For many aspects, this phase corresponds to the ``transition phase'' mentioned in \cite{2006MNRAS.368.1404A} for young, compact radio sources, until the forming cocoon reaches a well-defined shape (which may still evolve after that). Yet this transition requires longer times than predicted in that work (a few hundreds of thousands of years instead of a few tens of thousands) due to the complexity of the hydrodynamics.\\
 
  \item \textbf{``F'' phase: Forward propagation - }This stage shows quite complex hydrodynamics. Once the cocoon has formed, the internal jet propagates forward and may undergo some recollimation shocks, more likely for high $\mathcal{M}_{int}$. In the forward direction, the jet keeps building pressure in the HS. At the same time the jet is coupled to the cocoon, providing energy and gas to ``inflate'' it.\\ Asymmetry and irregularity in the cocoon start to be visible at this stage, together with significant backflows (see Section \ref{sec:backflow}); the contribution of the turbulent pressure $p_t$ is decreasing but still dynamically important (Figs. \ref{fig:presCocoon} and \ref{fig:thermo}). \\The cocoon axis ratio $r_{HS}/r_C$ seems here to settle on a well defined regime, but not necessarily self-similar. During this stage the jet pierces its own cocoon (see Section \ref{sec:geometry}), thus decoupling from it. This usually, but not always, coincides with the beginning of the next phase. Phase F lasts until about $2$ Myr (or 3, for enhanced density contrast runs), and corresponds to the second panel of Figure \ref{fig:evolution}. \\
\begin{figure*}
	\centering
    \includegraphics[width=0.249\textwidth]{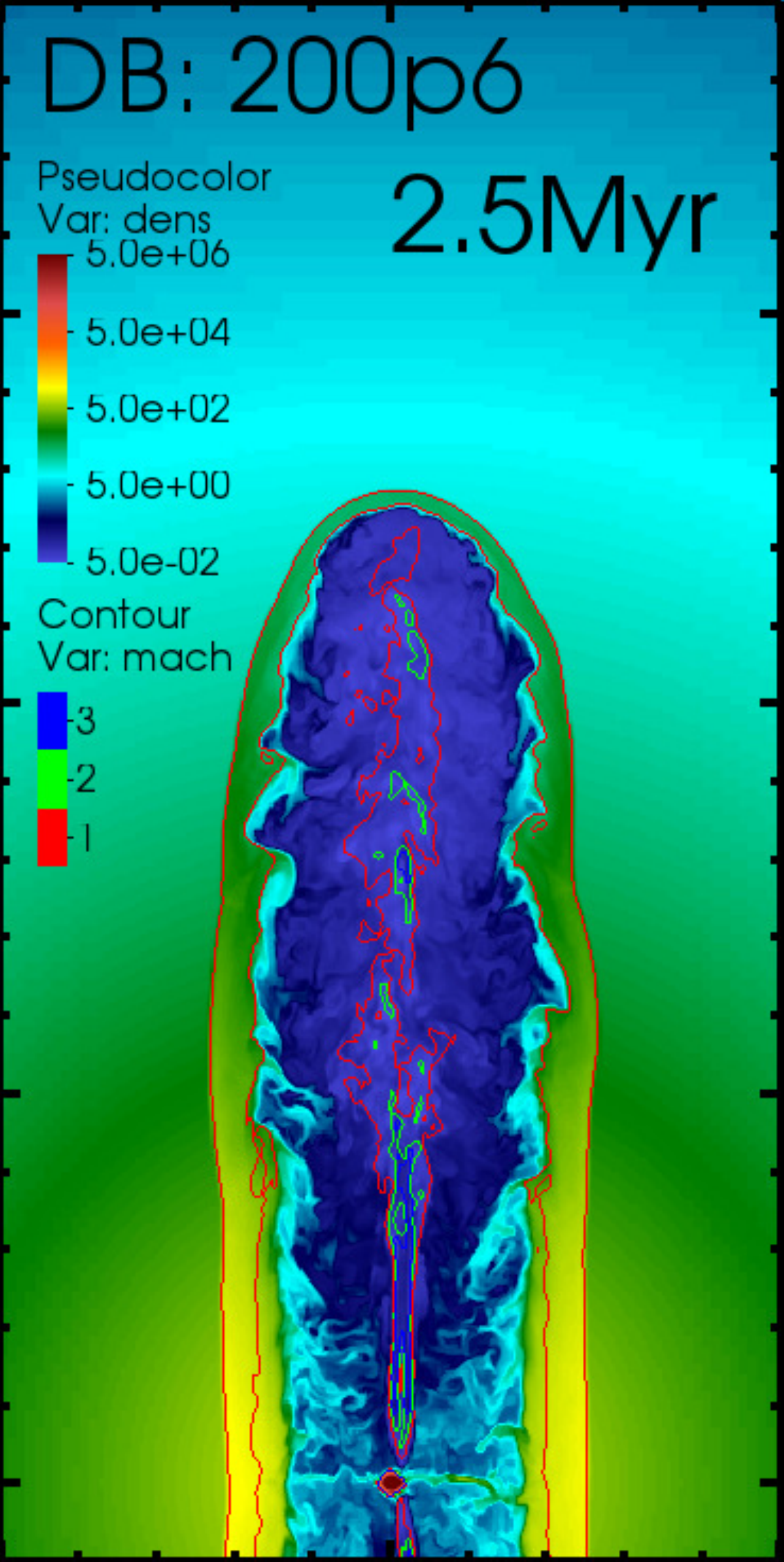}\includegraphics[width=0.249\textwidth]{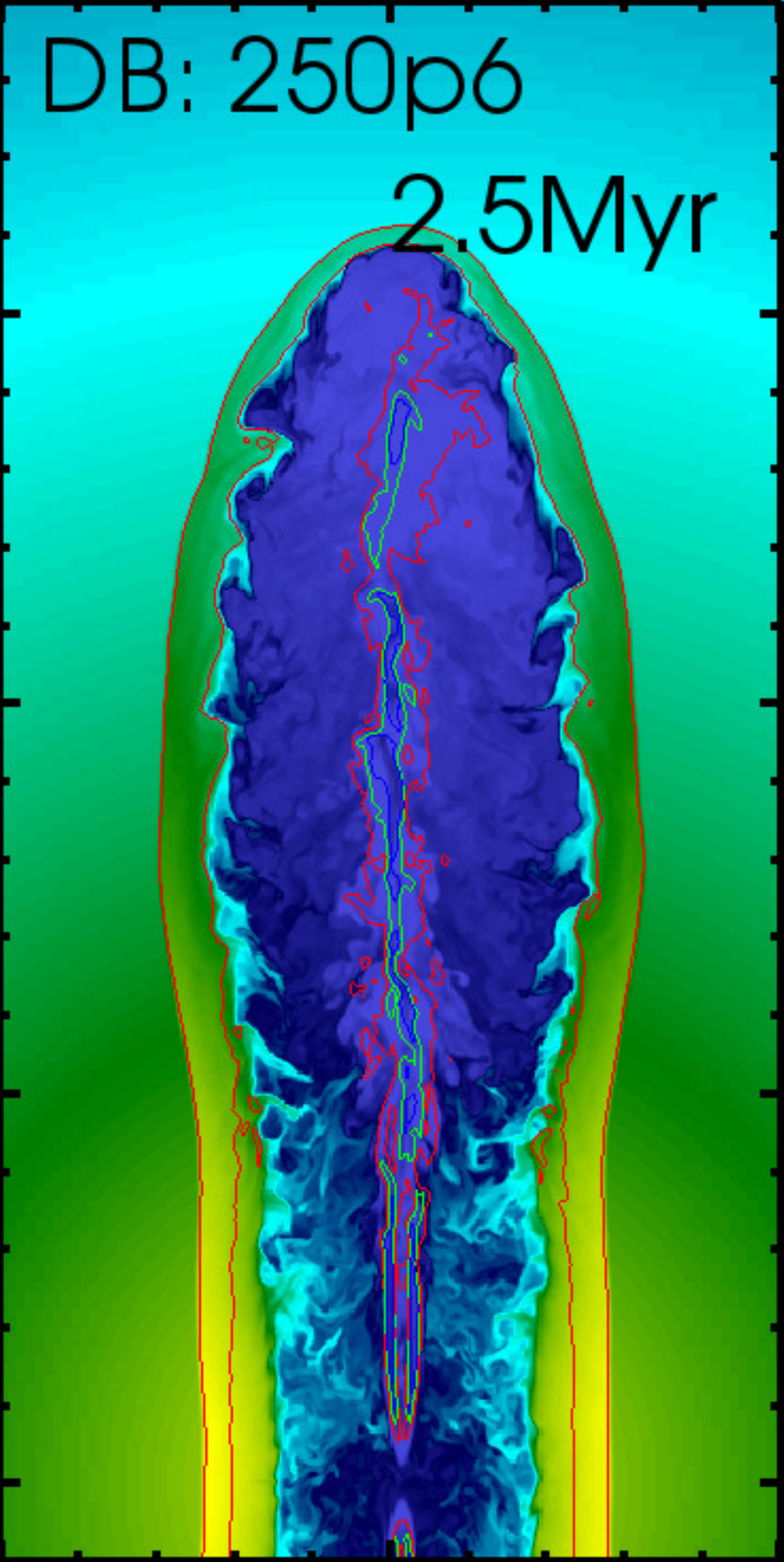}\includegraphics[width=0.249\textwidth]{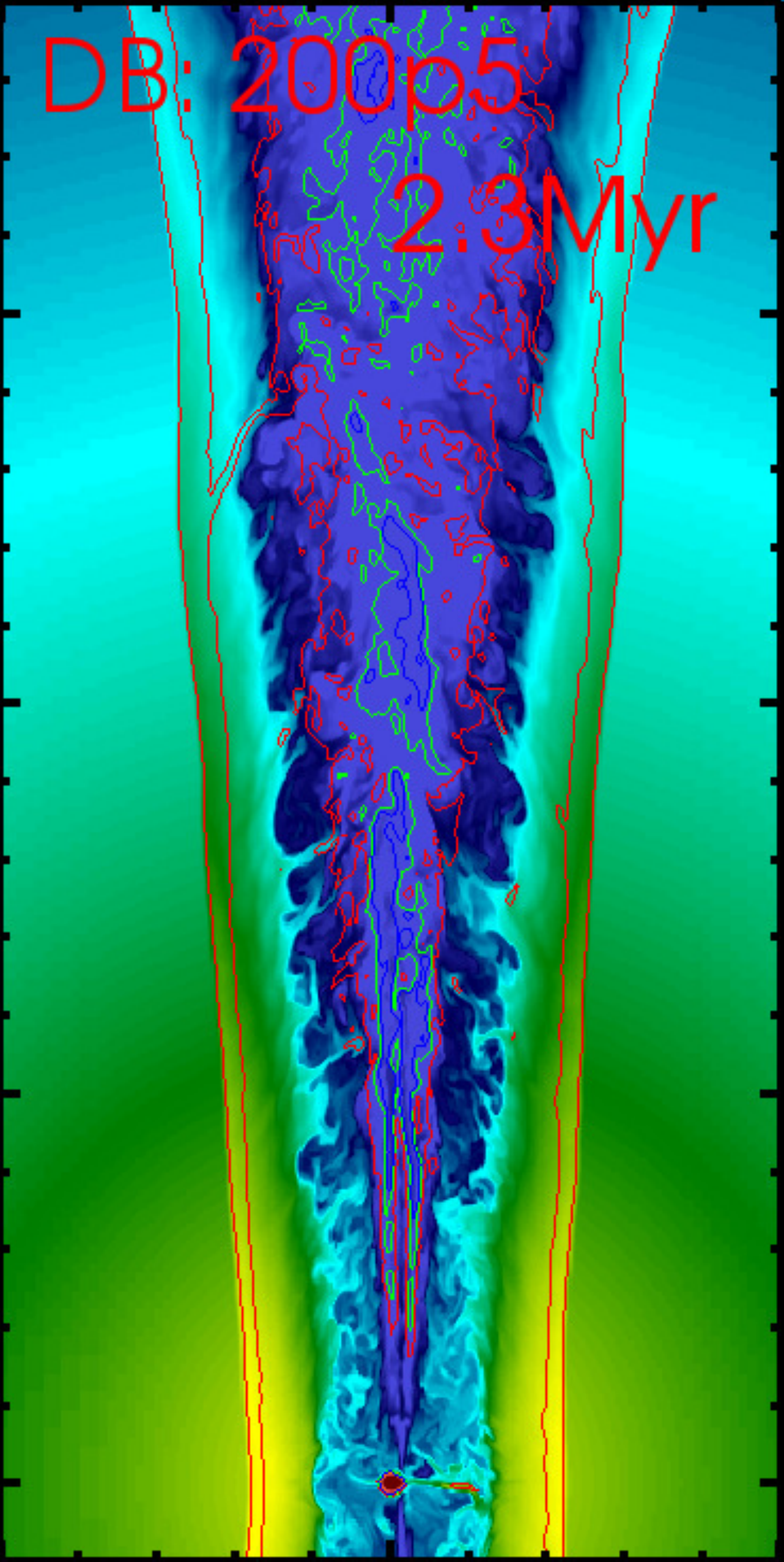}\includegraphics[width=0.249\textwidth]{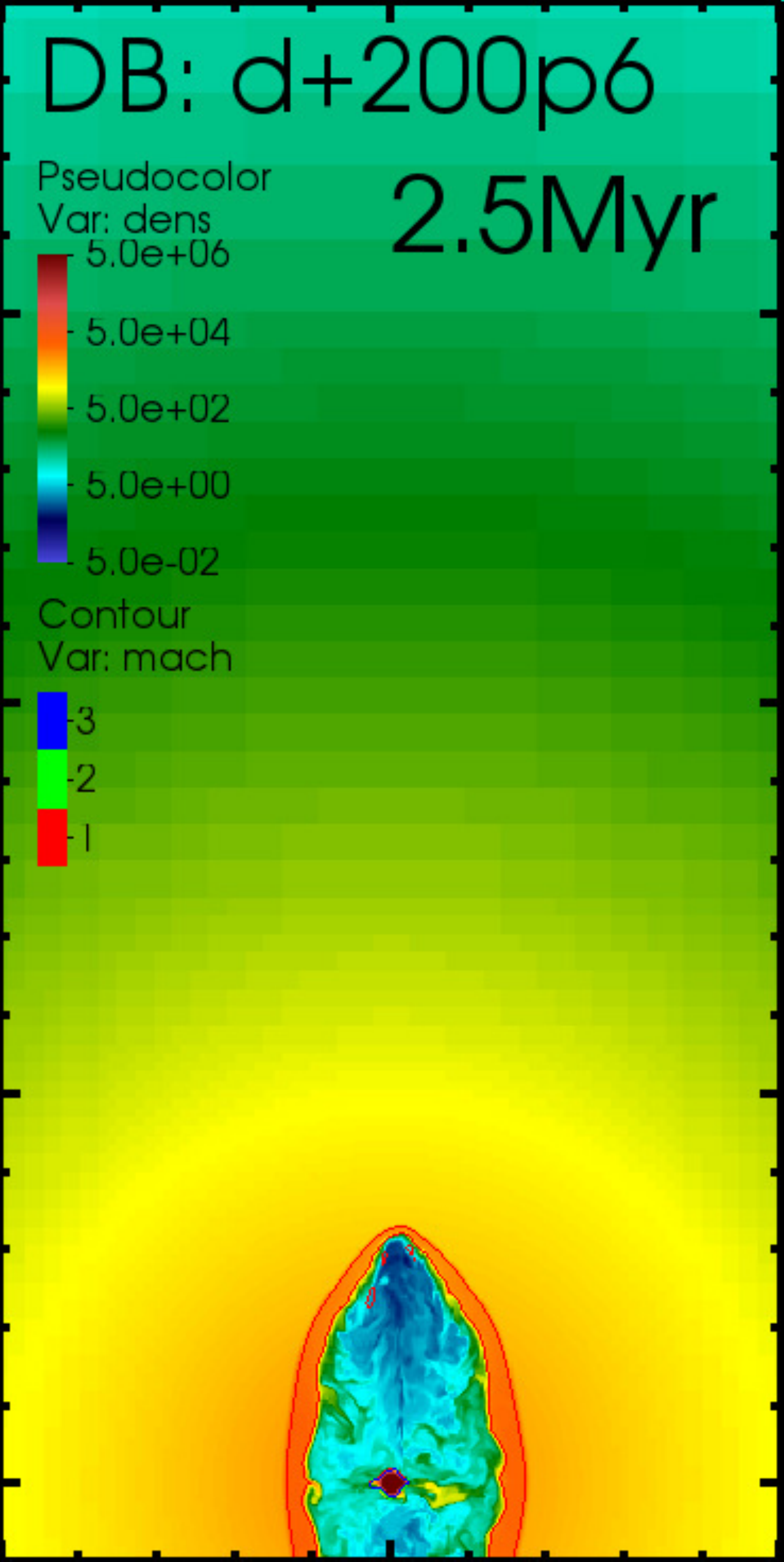}
	\caption{Visual comparison of different runs at $2.5$ Myr. Image size and color code are the same as Figure \ref{fig:evolution}. \textbf{First panel:} here we take run 200p6 as a reference model. In the next panels we change one parameter per time. \textbf{Second panel:}  run 250p6; increasing $\sigma_{V}$ leads to a faster and more penetrating jet. \textbf{Third panel:} run 200p5, shown at $2.3$ Myr only due to its very fast evolution. Decreasing $p_{jet}$, cocoon piercing events occur earlier, so the jets propagate faster, while the cocoon semi-minor axis $r_C$ remains smaller; lobes form faster. \textbf{Fourth panel:} run d+200p6 (note the higher ISM density) shows that an increased density contrast results in a rounder cocoon and a slower propagation.}
	\label{fig:visualCocoon}
\end{figure*}
  \item \textbf{``L'' phase: Lobe formation - }Right after the cocoon piercing, the ISM is no longer shielded from the jet; also, the ``naked'' HS is now in the outskirts of the halo (about $20$ kpc), where the gas density and pressure are not high enough for similar shielding effects. Thus, the gas coming from the jet and the inner part of the cocoon undergoes a fast and less directional expansion; the outcome of this is the formation of large lobes (a few tens of kpc, still expanding at the end of the simulation time) similar to the ones observed in FRII radiogalaxies, the HS being still well-defined (\citealp{2008ApJS..174...74K}).\\Due to this expansion, the cocoon semi-minor axis $r_C$ is now an ill-defined quantity (as there is no longer a cocoon); indeed in Figure \ref{fig:geoCocoon} it has a clear turn-up point. The denser ISM and light jets families runs show again a delayed behaviour, so not all of them were reached the state of having well-defined lobes; but all show cocoon piercing. The moment of piercing and the subsequent expansion are shown in the third and fourth panel of Figure \ref{fig:evolution}, respectively.
\end{enumerate}

Usually, after the cocoon piercing, the rapid expansion causes 	extsc{FLASH} to refine a very large volume, requiring much more memory; thus the simulation runs stop at this stage. In one case (200p5) the jet went out the simulation box before that happened.
The different components (jets, early lobes, cocoon material and bow shock region) are all highlighted in Figure \ref{fig:3Drender}.

This picture shows some elements in common with earlier simulations (\citealp{Sutherland:2007rz,2009MNRAS.396...61T,Gaibler:2010ij,2011ApJ...728...29W,Antonuccio-Delogu:2010ve}) and theoretical models (\citealp{1991MNRAS.250..581F,2006MNRAS.368.1404A}). Yet we are now able to link these elements with the internal dynamics, in a more organic picture. Details vary according to the different assumed ISM models, and the absence of cold gas in our runs. Comparison to 2D simulations performed with a similar setup (\citealp{2009MNRAS.396...61T}) shows striking differences. First, a much higher density contrast $\rho_{ISM}/\rho_{jet}$ is needed in order to recover the same cocoon shape, that otherwise is more extended in the forward direction; this could be due to 2D simulations not dealing properly with the turbulent pressure. The 2D simulations also tend easily to show too strong jet \emph{recollimation}.

The gas circulation inside the cocoon looks also very different (see Section \ref{sec:backflow}) once the third dimension is added.

\section{Cocoon geometry} \label{sec:geometry}
\begin{figure*}
	\centering
	\includegraphics[width=\textwidth]{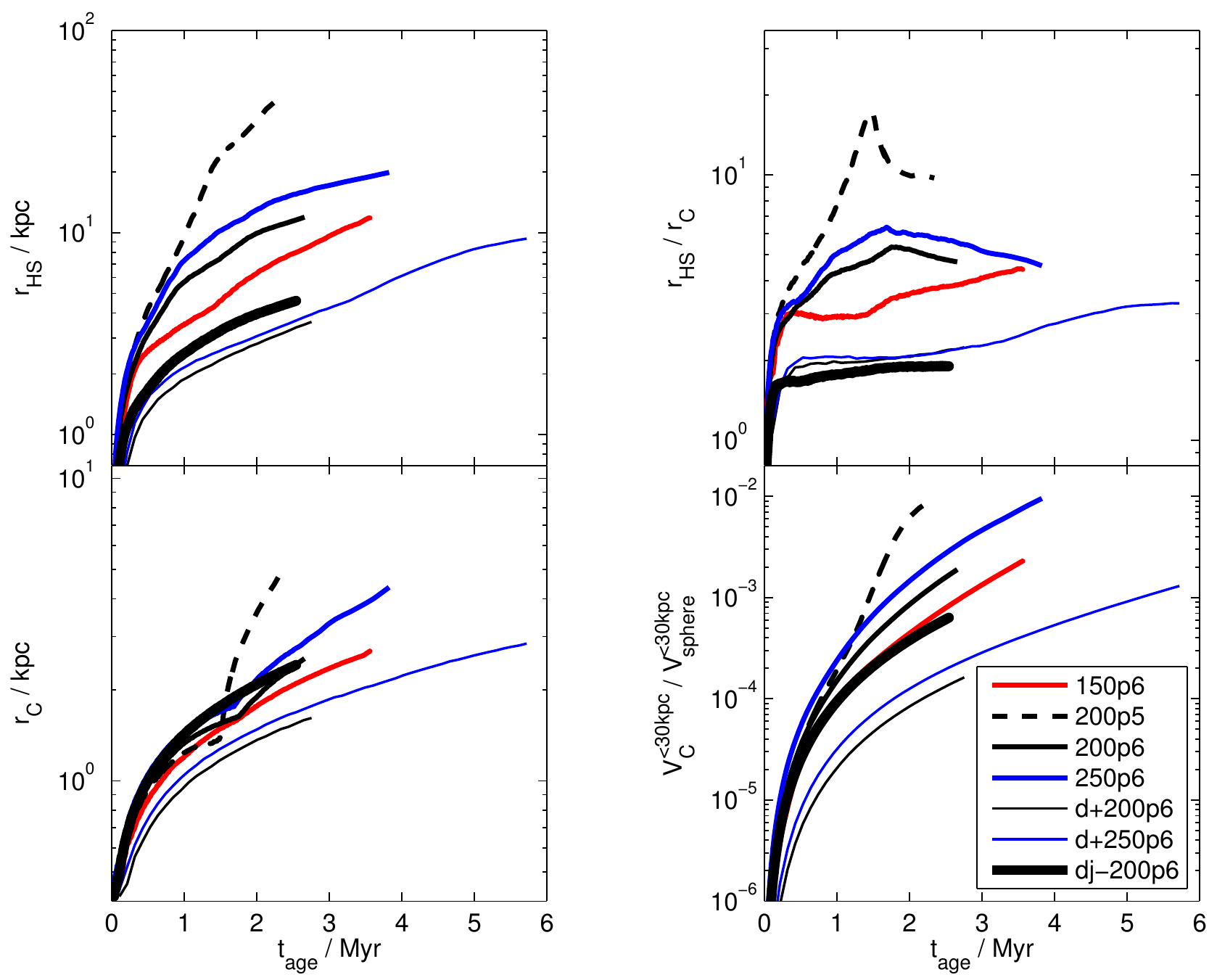}
	\caption{Geometrical properties of the cocoon (selected as the region where $T>2\times10^{9}$ K). \textbf{Plot style coding:} same line color means same $\sigma_{V}$, same line style means same $p_{jet}$, while line thickness discriminates \emph{fiducial} runs ($\rho_{jet} / \rho^c_{halo}$ = 0.01) from \emph{enhanced density contrast} runs ($\rho_{jet} / \rho^c_{halo}$ = 0.001). \textbf{Left:} time evolution of the cocoon average semi-major (top) and semi-minor (bottom) axes $r_{HS}$ and $r_C$. \textbf{Right: } cocoon aspect ratio $r_{HS}/r_C$ (up) and total volume $V_C$ inside a $30$ kpc-radius sphere(bottom) over time.}
	\label{fig:geoCocoon}
\end{figure*}
We want now to focus on the cocoon's shape and size in the different runs. What we had shown by visual impression in Figure \ref{fig:visualCocoon}, will be now quantified in Figure \ref{fig:geoCocoon}, where we consider the evolution of the cocoon's semimajor  and semiminor axes and volume.\\ These quantities have been calculated as follows: first we have selected all the cells in the mesh's grid in which the temperature is $\leq2\times10^9\rm{K}$;  this includes, for all runs, the cocoon region, the jets and the lobes, leaving out only the bow shock front and the unperturbed ISM. We refer to this operation as our ``cocoon selection''.

Then, the maximum extent along the jet axis define the semi-major axis $r_{HS}$ (top-left panel in Figure \ref{fig:geoCocoon}), while the maximum extent along the transverse direction defines the semi-minor axis $r_C$ (bottom-left panel). We also show the ratio $r_{HS}/r_C$ as a cocoon shape indicator (top-right); note that no shape constraint is assumed in the semimajor and semiminor axes extraction, which are two independent numbers.

As a last geometric property of the cocoon, we want to estimate the fraction of the surrounding ISM  that is affected by the jet. So we calculate the fraction of the cocoon volume $V_C$ with respect to a sphere centered on the jet origin, and having a radius of $30$ kpc (bottom-right). This volume fraction is then an indicator of the feedback activity on this scale. $V_C$ is just the sum of the volumes of the cells that pass the cocoon selection criterion, being another measure independent form the semimajor and semiminor axes.

In the first three panels of Figure \ref{fig:geoCocoon}, kinks in the curves are visible, clearly corresponding to phase changes. During phase C, i.e. the first $\sim0.3$ Myr the density contrast is the only parameter that plays an important role in determining $r_{HS}$ and the aspect ratio of the cocoon, so that the enhanced density contrast runs show the slowest forward propagation in favour of a less elongated cocoon shape. In other words, the cocoon inflation is a more isotropic process. 

After entering phase F, i.e. after $0.5$ to $1$ Myr, we can distinguish the effects of all the parameters. From the plots we see that $p_{jet}$ becomes the most important parameter; indeed the 200p5 (black dashed line) run shows little resistance from the ISM, and a very directional cocoon. The other fiducial runs decouple earlier from this trend, right after 1Myr or less, the earlier the lower $\sigma_V$. It is worth recalling that, due to the scaling relations used in our parametrization scheme, higher $\sigma_{V}$ means more massive haloes, but also higher jet power $P_{jet}$ and velocity $v_{jet}$. So, this simply means that more powerful jets propagate faster, provided that the injection pressure $p_{jet}$ is the same.

Nevertheless, for a density contrast of 1000, this is only a second order effect. Runs with this enhanced density contrast not only keep showing a less elongated cocoon which expands more slowly, but this expansion is also largely self-similar, with an aspect ratio close to the value of 2. Also, the aspect ratio can poorly distinguish  light jets and denser ISM runs,  the density contrast being more meaningful than the densities themselves (they in fact matter for $r_C$ and the total cocoon volume). This is seen in no other run, with the possible exception of run 150p6, in which phase F lasts too short a time to draw a conclusion. In general, the behavior of the fiducial runs in this phase is quite complex and difficult to interpret, suggesting a very strong dependence on internal dynamics.

The cocoon geometry is  well captured by the jet injection Mach number $\mathcal{M}_{int}$ (Table \ref{tab:run}); the lower it is, the slower and less elongated the cocoon will be. Cocoons created by jets with the same $\mathcal{M}_{int}$, will be more spherical if the density contrast is higher.
\begin{figure*}
	\centering
    \includegraphics[width=1\textwidth]{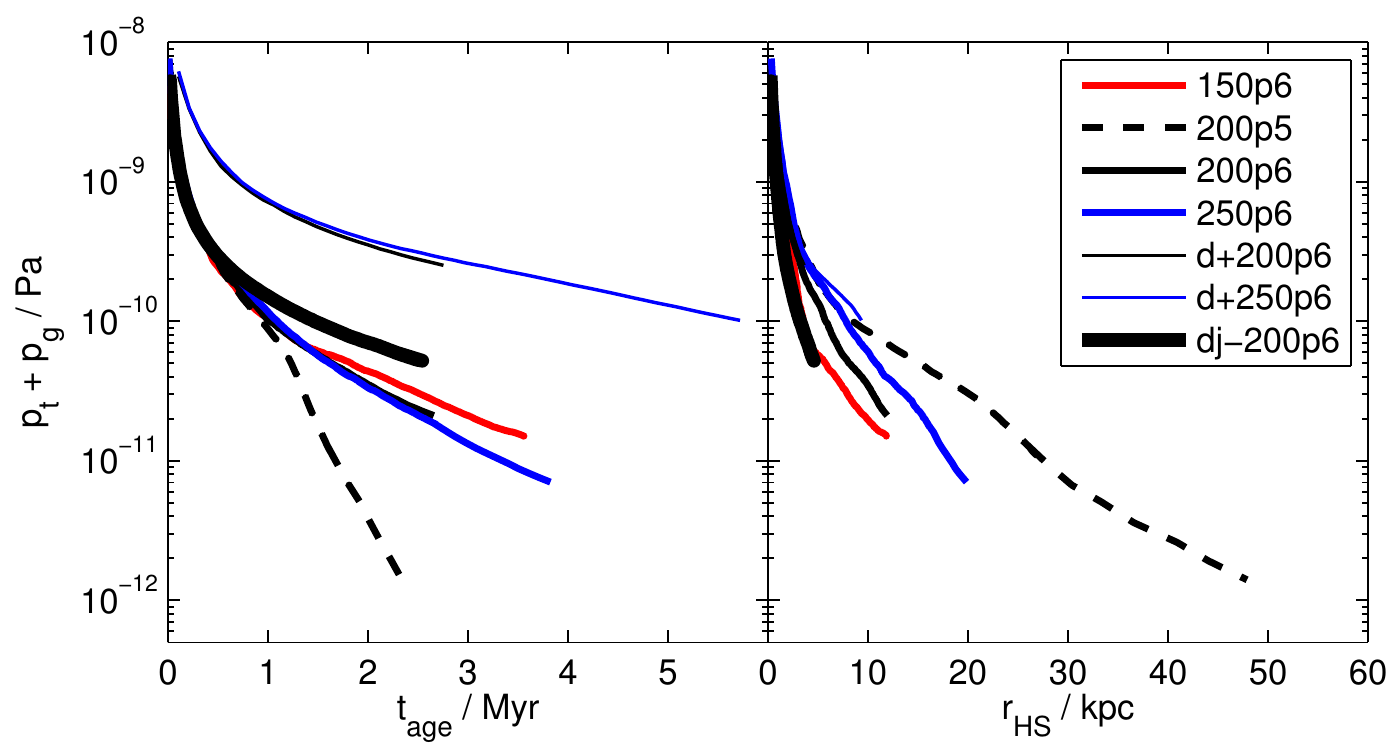}
    \caption{Total pressure (gas + turbulent) averaged inside the cocoon ($T>2\times10^{9}$ K), as a function of time (left) and of cocoon semi-major axis $r_{HS}$. The turbulent pressure is computed by Reynolds' decomposition (see text). \textbf{Plot style coding:} same as Figure \ref{fig:geoCocoon}.}
	\label{fig:presCocoon}
\end{figure*}
The moment of cocoon piercing, when phase L is entered, is clearly marked by an upwards kink in the aspect ratio and even more in $r_C$, which now measures the lobe transverse radius rather than the cocoon's. In turn, $r_{HS}$ is often little affected. This explains why in a Fanaroff-Riley type II galaxy the jet will always be confined by a HS at its end\footnote{This is sometimes referred to as the jet never turning ``ballistic''} (and a HS is always present in our simulations too). So $r_{HS}/r_C$ gets lower in most simulations due to the lobe expansion. In runs such as d+250p6 (thin blue solid line) yet this ratio increases. This behavior is illustrated in Figure \ref{fig:evolution}: here cocoon piercing does not yet start the lobe creation, but the system is still in its F phase. In order to understand this, we provide some more insight on the piercing mechanism. The bow shock region has both an inner and an outer front, both expanding at the local sound speed (the red contours in the figure mark where the local Mach number is equal to 1). The layer comprised between these two fronts is yet very thin (just a few cells thick, say $\leq50$pc) near the HS. In an elongated cocoon (like in the fiducial runs) this thin edge will be completely destroyed, giving a large opening for the jet to come out and expand in the lobes. In a more spherical cocoon the layer will be this thin only, say, within a few hundred parsecs from the HS; this configuration will thus offer a smaller hole to the jet. The lobes' expansion is likely to occur at later times for these runs.

The cocoon volume (bottom right panel) increases smoothly with no clear sign of phase change, but it never exceeds $1\%$ of the volume of a sphere of 30kpc radius; thus the ISM fraction affected by the jets is hardly significant in this stage. So feedback from \emph{early} AGNs jet may have only very limited impact in the host galaxy. The cocoons and lobes are in fact still expanding, so one may be confident that in fully-developed sources the scenario will be different.

\section{Cocoon thermodynamics} \label{sec:thermodynamics}

We have performed an analysis of the thermodynamic state of the cocoon; we show its temperature and pressure, together with the energy it  exchanges with the external ISM. 
\begin{figure}
	\centering	
    \includegraphics[width=1\columnwidth]{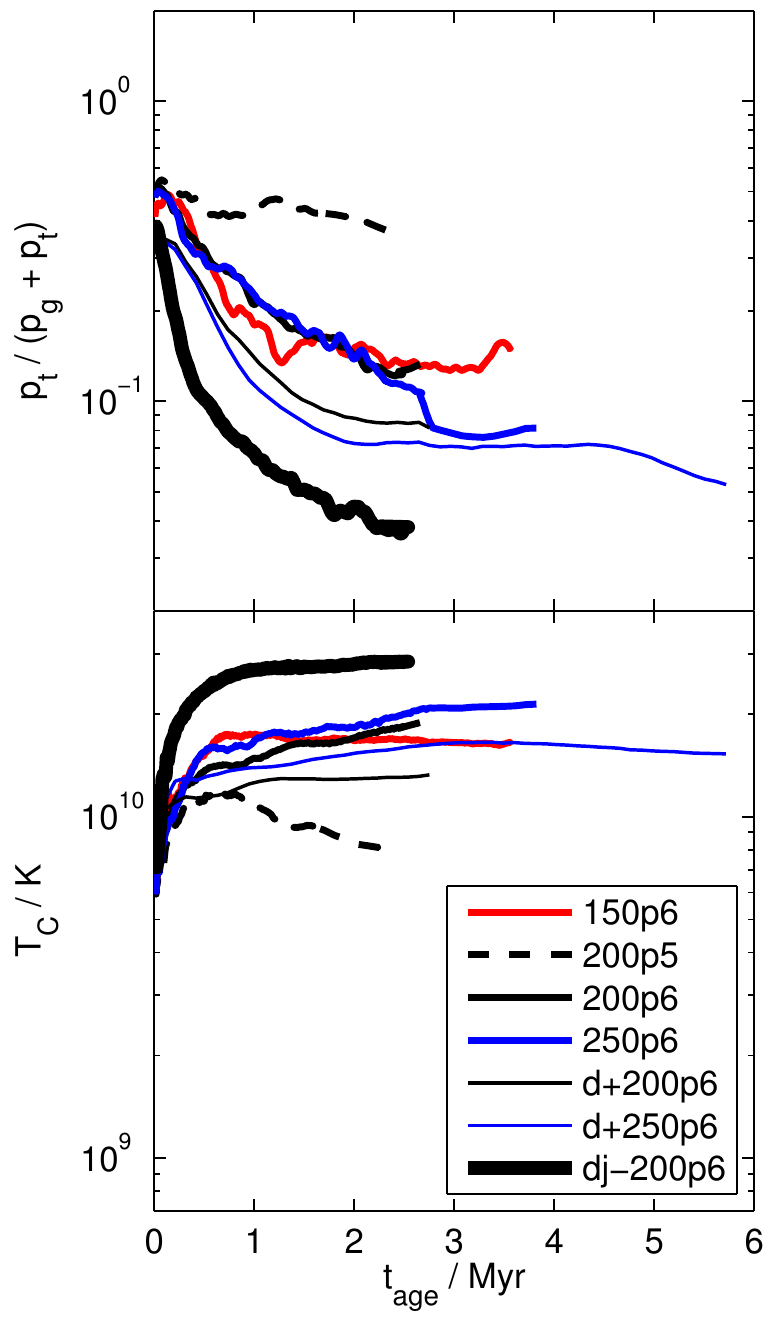}
    \caption{Thermodynamics of the cocoon ($T>2\times10^{9}$ K). \textbf{Plot style coding:} same color means same $\sigma_{V}$, same line style means same $p_{jet}$, while line thickness discriminates \emph{fiducial} runs (${\rho_{jet}}/{\rho^c_{halo}}=0.01$) from \emph{enhanced density contrast} runs ($\rho_{jet} / \rho^c_{halo}$ = 0.001). \textbf{Upper panel:} Evolution of the turbulent fraction of the cocoon pressure ${p_t}/{(p_t+p_g)}$. \textbf{Lower panel:} Average cocoon temperature $T_C$ as a function of time.}
	\label{fig:thermo}
\end{figure}
In the following, whenever we refer to an \emph{intrinsic} quantity, we mean its average value within the cocoon; this does not necessarily imply that the cocoon is in a thermodynamic equilibrium state. In turn, this \emph{cocoon average} is always weighted by the corresponding \emph{extrinsic} quantity; so that every \emph{cocoon average} will be always operated on quantities having the physical dimension of an energy. For instance, pressures will be weighted by cell volumes; velocities (which appear in the turbulent pressure calculation) will be squared and weighted by cell masses, etc.

In Figure \ref{fig:presCocoon} we show the cocoon mean \emph{total pressure}, as a function of both $t_{age}$ and $r_{HS}$ (first and second panel, respectively). This pressure is defined as the sum of the gas (hydrodynamic) pressure $p_g$ and the turbulent pressure $p_t$. The latter has been calculated through a standard Reynolds decomposition (the trace of the Reynolds' tensor), assuming as unperturbed velocity for each cell the mean velocity of its parent block. This is a natural choice, following directly from the AMR structure of our simulation: blocks and cells have different sizes according to their specific refinement level; so the scale on which we have to study (or we can resolve) turbulent motions varies in the same way.

When the ISM is denser (thin solid lines), the cocoon pressure as a function of time is much larger (roughly by a factor of 10, still increasing after the first 2Myr), but this is just because  the pressure of the external ISM is likewise increased by a factor of 10 with respect to the fiducial cases. It is interesting to notice how this difference disappears in the second panel: cocoons with higher pressure will expand more slowly (see Figure \ref{fig:geoCocoon}) and thus (partially) compensate for this difference. For the same reason, models such as \cite{1997MNRAS.286..215K} state their predictions for the pressure as a function of $r_{HS}$ rather than $t_{age}$. Direct, quantitative comparison with these predictions would yet be of little significance and difficult  to interpret, because of the different assumptions about the ISM density distribution.

Besides the trivial aforementioned density differences, all the curves decrease smoothly up to phase F; later, the rapid cocoon expansion in the 200p5 (dashed black line) runs makes its pressure turn down by two order of magnitudes in about 2Myr; much faster than the other ones. Density contrast and $\sigma_V$ also play an important role, in concordance with the geometric properties described in the previous section. The general picture that we deduce is that a jet capable -for any reason- of building up a higher pressure, will result in a less elongated cocoon: the pressure, as expected, promotes isotropic expansion. In turn, a cocoon (the bow shock front, to be more precise) with a lower $r_{HS}/r_C$ will need to move more gas from the ISM in order to inflate; thus it will expand more slowly.

The turbulent pressure fraction ${p_t}/{(p_t+p_g)}$ is shown in Figure \ref{fig:thermo} (upper panel). During the first few tens of thousands years, $p_t$ is very close to 25$\rm{\%}$ of the total, in all the runs. Run 200p5 is again an extreme case, never showing signs of decay from this value. This can be explained in the following terms: for the consideration we expressed in Section \ref{sec:setup}, a lower $p_{jet}$ means a higher $v_{jet}$, so a higher shear inside the cocoon (compare runs in Figure \ref{fig:visualCocoon}), thus we can expect more turbulent motions. The fiducial runs (lines of intermediate thickness), are indistinguishable until $2.5$ Myr. The \emph{denser ISM} family converges to a little less than 10$\rm{\%}$ after $\sim2$ Myr; in the light jets run the $p_t$ fraction is already below 3$\rm{\%}$ at that time. Almost all runs show, if not yet in complete convergence, that  stability develops around these values.

The lower panel in \ref{fig:thermo} shows the time evolution of the average cocoon temperature $T_C$. Again, after a transition corresponding to phase C, all runs (except for 200p5) converge to some value in the range $\left[1,\,3\right]\times10^{10}$ K, so that the jet-powered expansion is to a considerable extent an isothermal process. The actual convergence temperatures just reflect the thermodynamic state of the jet at injection.
This convergence in both $T_C$ and the $p_t$ fraction means that some \emph{self-regulation mechanism} is at work, stabilizing the turbulent pressure. Yet, this happens regardless of whether self-similarity in the expansion is achieved or not, while runs such as 200p5, undergoing fast expansion and thus relatively little interplay with the ISM, do not show such a convergence. So self-regulation appears more related to the interaction with the ISM, and all the complicated hydrodynamics therein (cocoon piercing, recollimation shocks, backflows; see Section \ref{sec:backflow}), rather  than to the geometry of the expansion, to which it is more often linked.

\begin{figure}
	\centering
	\includegraphics[width=0.95\columnwidth]{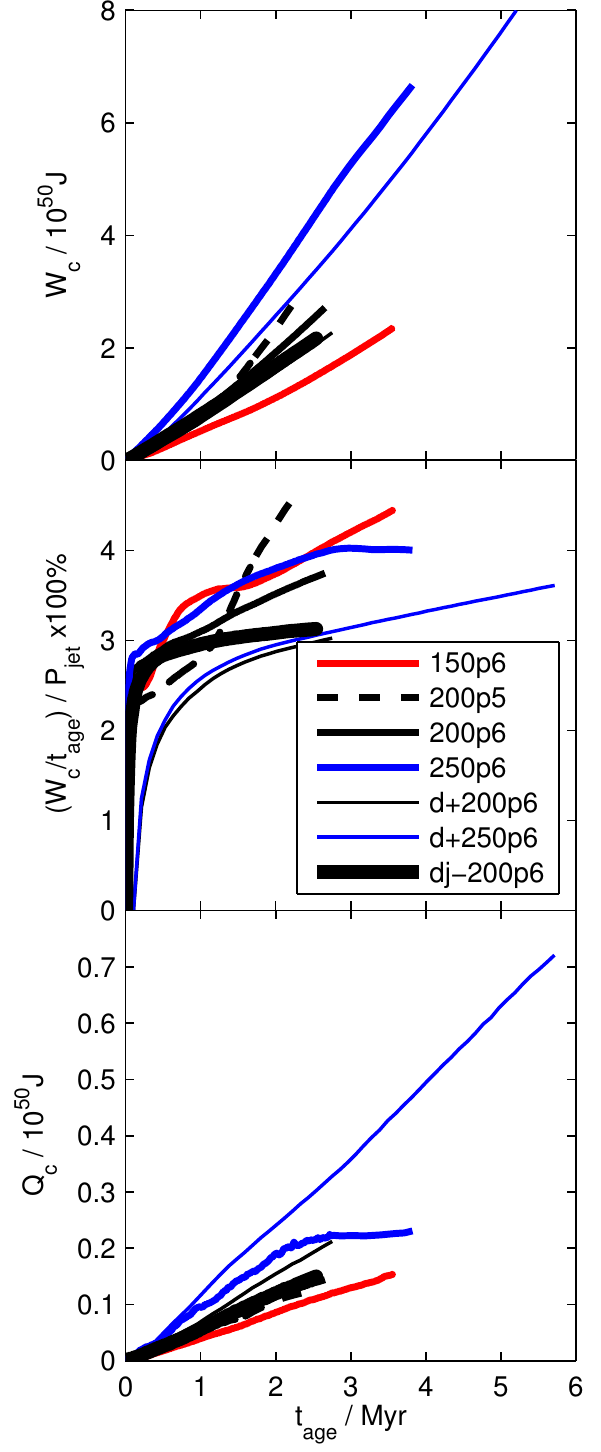}
	\caption{Mechanical \emph{$pdV$} work $W_{C}$ done by the cocoon expansion on the rest of the ISM (upper panel), (time-averaged) power developed by $W_C$ as a percentage of the input power $P_{jet}$ (middle panel) and \emph{$TdS$} heat exchange $Q_{C}$ as a function of time. The energies $W_C$ and $Q_C$ quantities are calculated as cumulative integrals in the cocoon volume-pressure and entropy-temperature diagrams, respectively. All values suggest that jet feedback is energetically significant even in the first Myr. \textbf{Plot style coding:} same as Figure \ref{fig:presCocoon}.}
	\label{fig:energy}
\end{figure}

Finally, in Figure \ref{fig:energy} we plot the energy exchanged between the cocoon and the ISM: the upper panel shows the \emph{cumulative} $pdV$ work of the cocoon $W_C$, while the lower panel likewise contains the $TdS$ exchange of heat $Q_C$. The entropy $S$, here and in the following, is calculated as in \cite{Tooper69:entr}, by:
\begin{equation*}
	S = \rho\frac{N_{Av}k_{B}}{\mu}\,\ln\left(\frac{T^{1.5}}{\rho}\right)
\end{equation*}
where $\mu=0.5988$ is the mean molecular weight, while the temperature $T$ and the density $\rho$ are evaluated in each cell.
Both quantities increase nearly linearly during phases F and L, so that the energy deposition, i.e. the essence of the feedback,  is constant with time. The mechanical work $W_C$, that \cite{Antonuccio-Delogu:2007oq,2009MNRAS.396...61T} associate to gas compression and ultimately to positive feedback, constantly outnumbers by an order of magnitude the exchanged heat $Q_C$ (calculated as integral in the temperature/entropy state diagram), associated to the negative feedback. This also points to an early positive feedback in the innermost kpc (\citealp{2009MNRAS.396...61T}, \citealp{Gaibler:2011hc}). Yet, the presence of cold gas and thermal conductivity may change this value significantly, so this must be taken as a lower limit.

Finally, the middle panel shows the ratio $\left(W_C/t_{age}\right)$ as a percentage of the injection power $P_{jet}$ ($P_{jet}$ is reported in Table \ref{tab:run}); in other words, this quantity is the time average up to the instant $t_{age}$ of the \emph{jet/ISM energy coupling constant}.  We notice that in these first few Myr,  this is always within $\left[3,\,5\right]\%$; such values of coupling  are believed to be very significant in the  galaxy formation context (e.g. \citealp{2007MNRAS.380..877S,2013MNRAS.432.3381M}).

\section{Backflow} \label{sec:backflow}
\subsection{Overview on Backflow} \label{sub:backflowOverview}

Within the jet-cocoon system we can distinguish few flow structures with different (and time-varying) levels of regularity. Motion within the jet is mostly laminar, but Kelvin-Helmoholtz instabilities along its path tend to produce turbulent eddies and may destabilize this flow. These turbulent eddies propagate within the cocoon and result in transonic turbulence. 
\begin{figure*}
  \centering
    \includegraphics[width=0.33\textwidth]{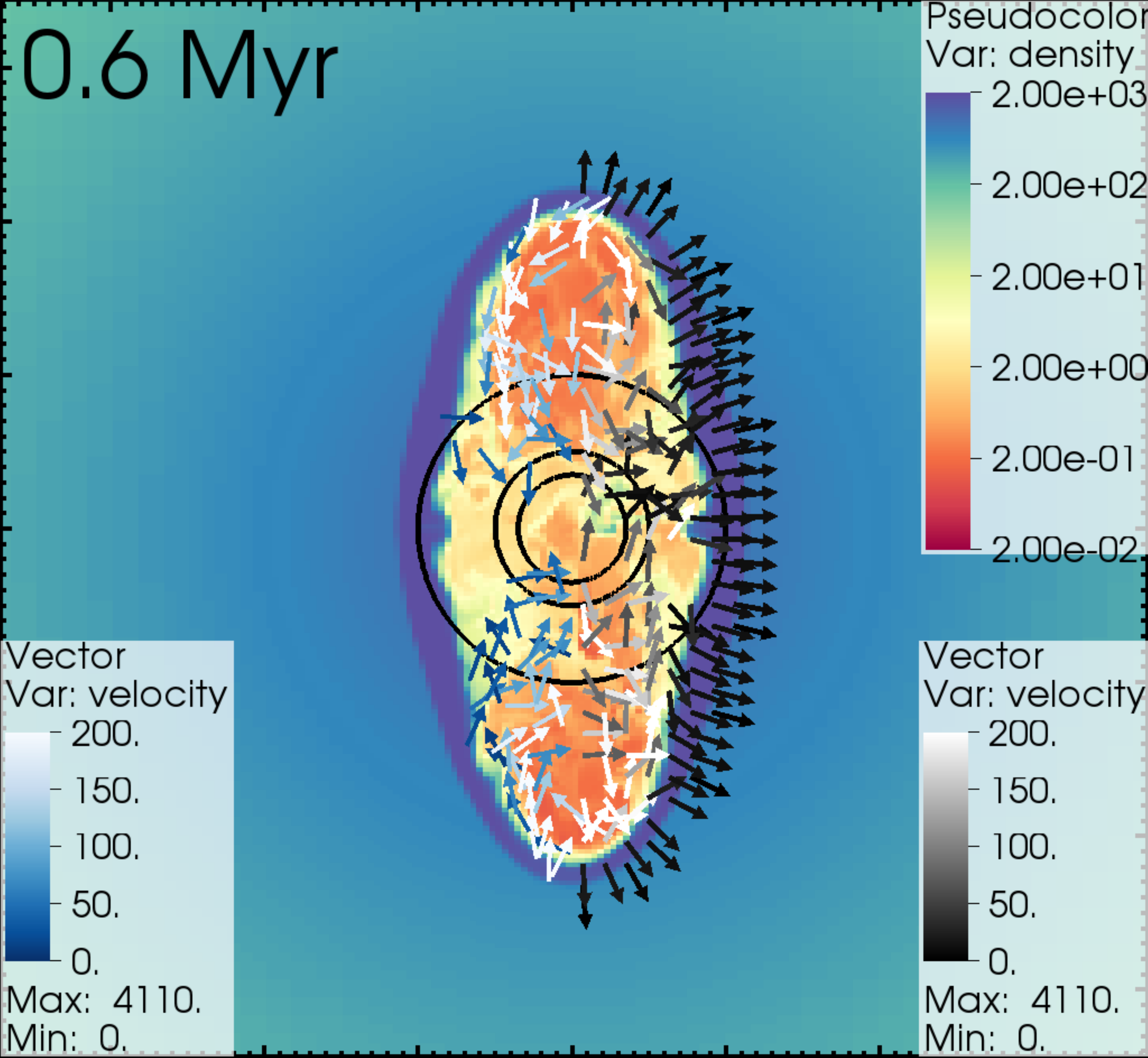}\includegraphics[width=0.33\textwidth]{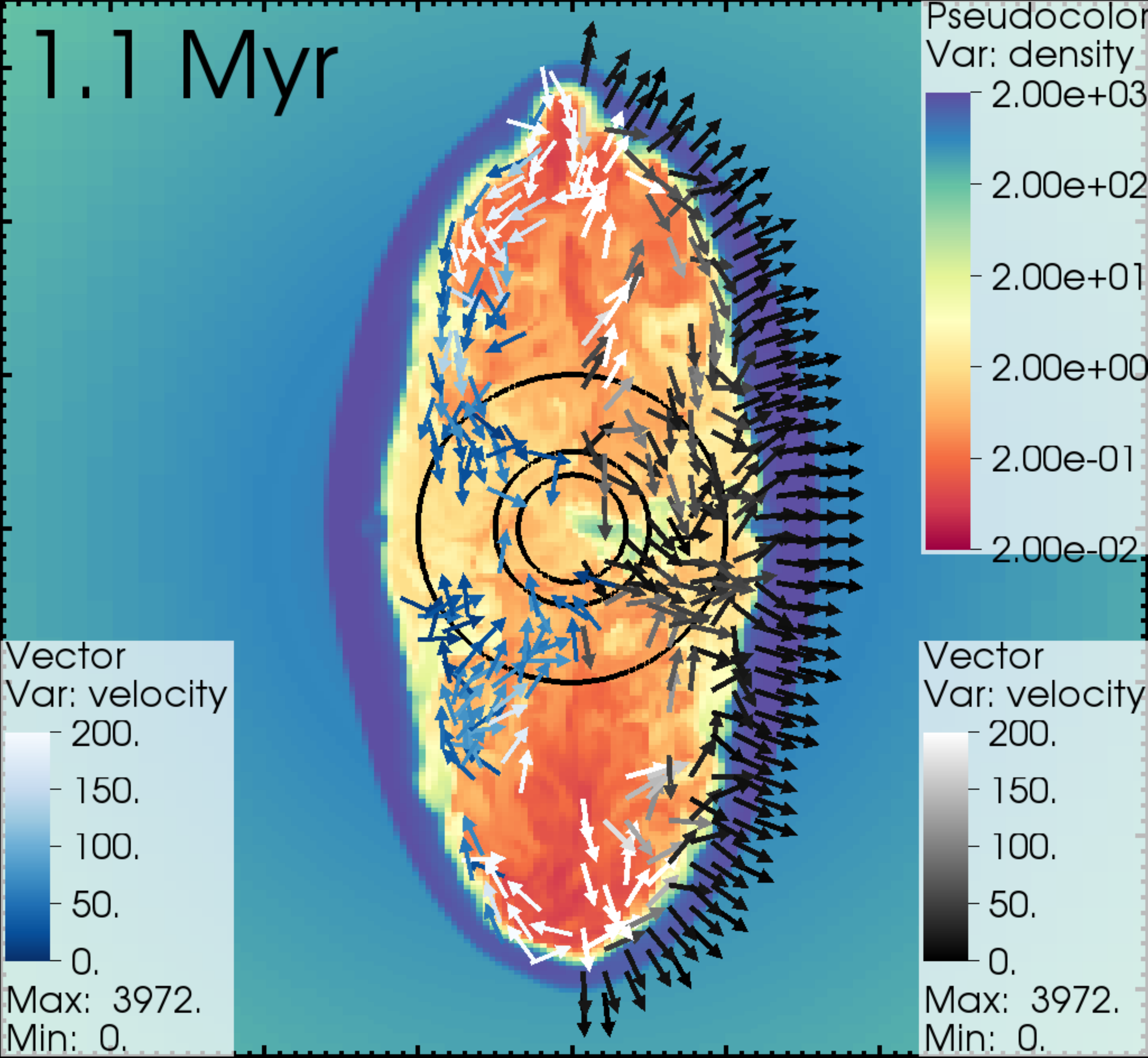}\includegraphics[width=0.33\textwidth]{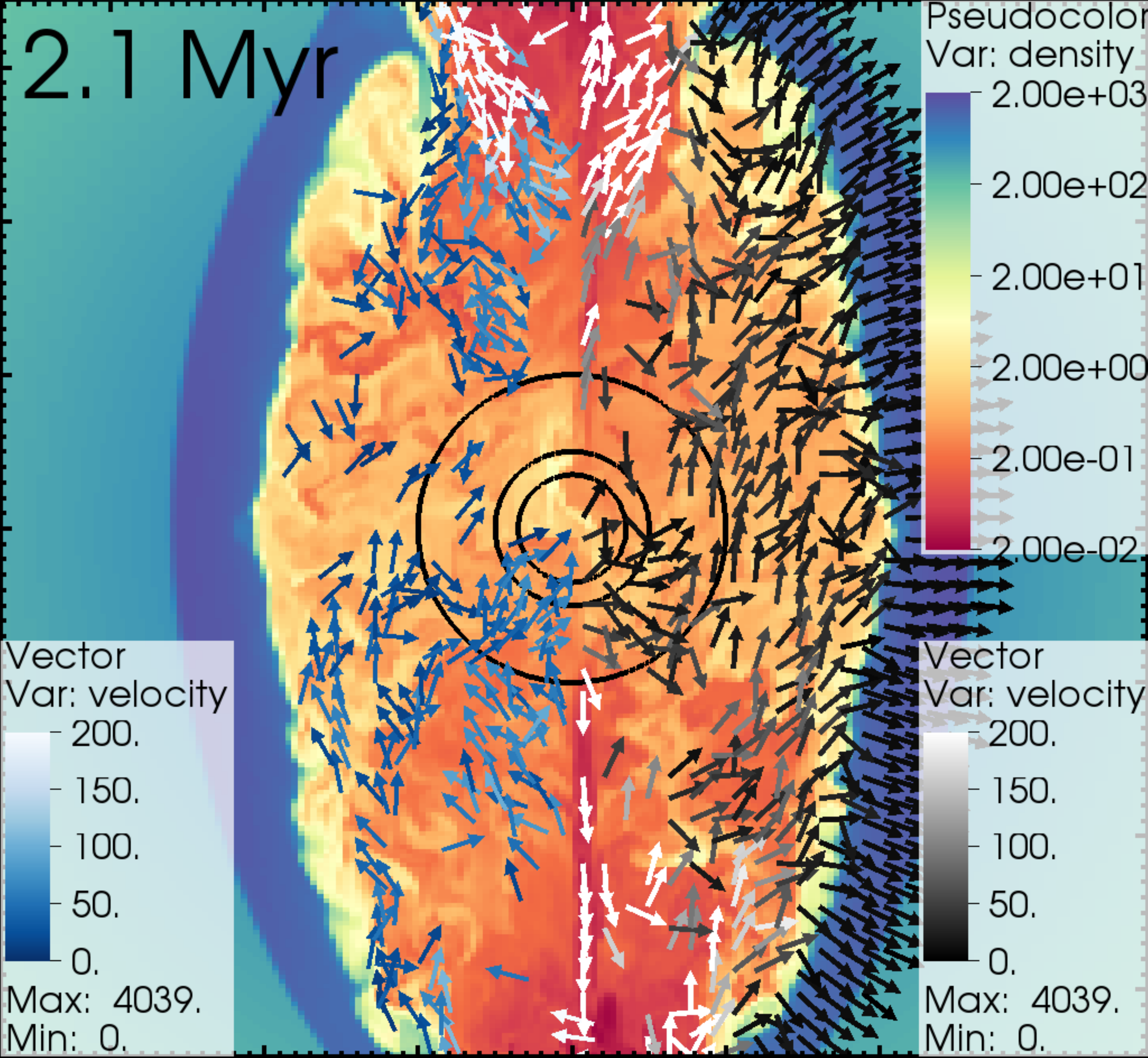}
    \caption{Central slices of run dj-250p6, at $t_{age}=0.6$, $1.1$ and $2.1$ Myr. Box size: $7$ by $7$ kpc. The background shows the gas density (pseudocolor plot). The arrows trace the velocity field $\mathbf{v}$. Left half (blue to white): backflow velocity only, i.e. $\mathbf{v}\cdot\hat{z}<0$. Right half (black to white): projected expansion velocity on the x-y plane (i.e. where $xv_{x}+yv_{y}>0$).	Contours at $0.35$, $0.5$, and $1$ kpc distance from the centre are also shown as black circles.}
    \label{fig:velField}
\end{figure*}

Finally, we also observe a \emph{backflow} within the cocoon. By this term we define a \emph{spatially coherent} flow directed opposite to the jet itself. In all the runs we have performed, this backflow develops during the initial $10^{5}$ - $10^{6}$ years. This feature of jet-cocoon systems was already noticed in the first simulations of the propagation of relativistic jets into homogeneous atmospheres \citep{1982A&A...113..285N}, and confirmed by more recent simulations \citep{2008A&A...488..795R,2007MNRAS.382..526P}.

\citet{2010ApJ...709L..83M} distinguish backflows, according to their morphologies: a \emph{straight} backflow, with flow lines extending from the tip of the hotspot back to the origin, and a \emph{bent} one, where the flow lines are instead bent near the meridional plane. In their previous 2D simulations, \citet{2010MNRAS.405.1303A} also noticed the formation of these features, and noticed that the backflow turned from a 
\emph{bent} to a \emph{straight} morphology with evolving time.

\begin{figure}
  \centering
    \includegraphics[width=0.95\columnwidth]{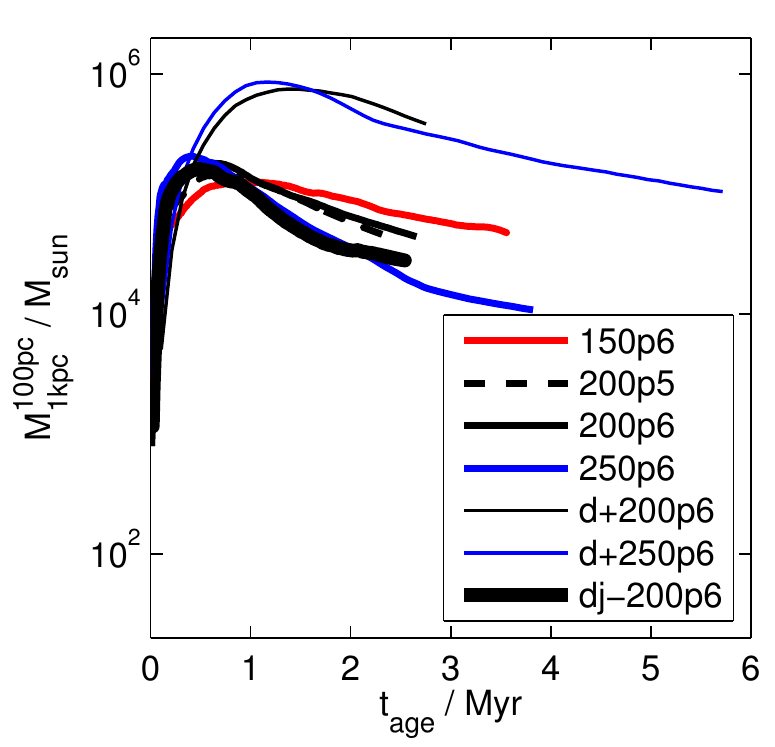}
    \caption{Total mass evolution around the central region of the jet-cocoon system. We compute the total mass within $1$ kpc from the centre, excluding the central $0.1$ kpc to avoid possible numerical contamination. We use our standard cocoon selection ($T\geq2\times10^{9}$ K) plus a density threshold ($\rho\leq4.23\times10^{-1}\rm{ {cm}^{-1}}$). In this way we eliminate from the computation both the central overdense region and the hot material from the jet. In the \emph{fiducial} and the \emph{light jets} runs, the increase in total mass during the first few $10^{5}$\,yr is mostly contributed by the backflowing gas converging towards the meridional plane. The same is true for runs from the \emph{denser ISM} family (thin lines), but the mass increase is higher and peaks at later times. This is just due to the larger mass in the central region. \textbf{Plot style coding:} same as Figure \ref{fig:geoCocoon}.}
  \label{fig:mass-flow}
\end{figure}

As shown in Figure~\ref{fig:velField}, a \emph{straight} backflow arises during the early evolution of the cocoon (phase C and early phase F), spatially confined between the jet and the bow shock. Until about $0.6$ Myr the backflow is coherently organised along streamlines extending almost continuously from the hotspots to the meridional plane, thus contributing to replenish the central accretion region and the disc with gas. However, the turbulence which develops within the cocoon acts to heavily perturb these backflows, and their large-scale coherence is completely destroyed after $2-3\,$ Myr. In order to understand how much this backflow can affect the central SMBH, we plot in Figure \ref{fig:mass-flow} the evolution of the total gas mass and within a sphere of $1$ kpc radius, centered at the origin of the jet. We also remove the innermost $100$ pc, that may introduce numerical error contamination (but gas accreted in the innermost kpc likely ends up in that region).

The backflow region is selected, besides the standard cocoon selection ($T\geq2\times10^9$ K), also by a density threshold criterion ($\rho\leq4.23\times10^{-1}\rm{ {cm}^{-1}}$), in order to select gas contributed from the backflow \emph{but} not belonging to the jet. In all but the \emph{denser ISM} runs, the mass accumulates from the innermost regions and proceeds towards the external regions, on time scales of $t_{acc}\sim10^{5}$ years, followed by a slower decrease ($t_{age}\sim2-3$ Myr). 

A maximum mass of $0.8-2\sim10^{5}\, \rm{M_{\odot}}$ is accreted by the backflow, without any appreciable dependence of the time-scales of the backflow on either $P_{j}$ or on $\sigma_{v}$, (i.e. on the global galaxy mass) for the \emph{fiducial} runs. For those runs where the central density is 10 times larger (thin lines),  we observe  a similar behavior, except that  the global mass of the backflow is correspondingly 10 times larger, and the decay time is longer ($\simeq1$ Myr). Thus, we conclude that the ISM density is the only parameter which determines the amount of gas which the backflow can drive back towards the SMBH accretion region.

If we compare our backflows with the ones in \cite{Antonuccio-Delogu:2007oq}, we find that ours drive more gas to the central region during the first few $t_{acc}$; later, they are still present, but only occasionally they reach the central region; indeed they propagate with ease for $\aplt15$ kpc from the HS ($\aplt5$ kpc for the enhanced density contrast runs), but fade after that distance. 
\begin{figure}
  \centering
    \includegraphics[width=1\columnwidth]{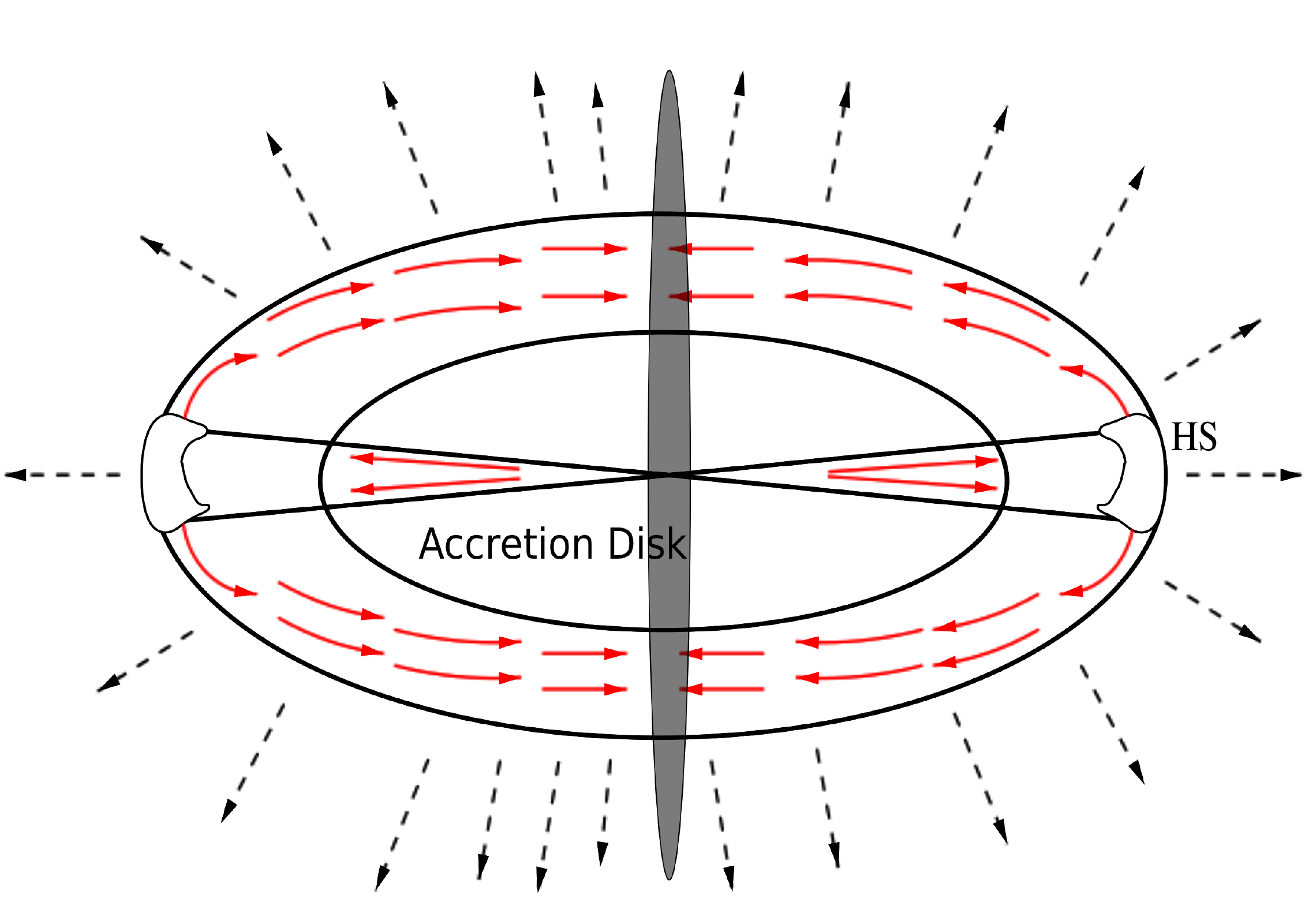}  
    \caption{Model of backflow. As the cocoon expands laterally, the outermost backflow streamlines are driven farther away and the material which they carry reaches the disc at progressively later times.}
  \label{fig:mass-flow_model}
\end{figure}

\begin{figure}
  \centering
    \includegraphics[width=0.95\columnwidth]{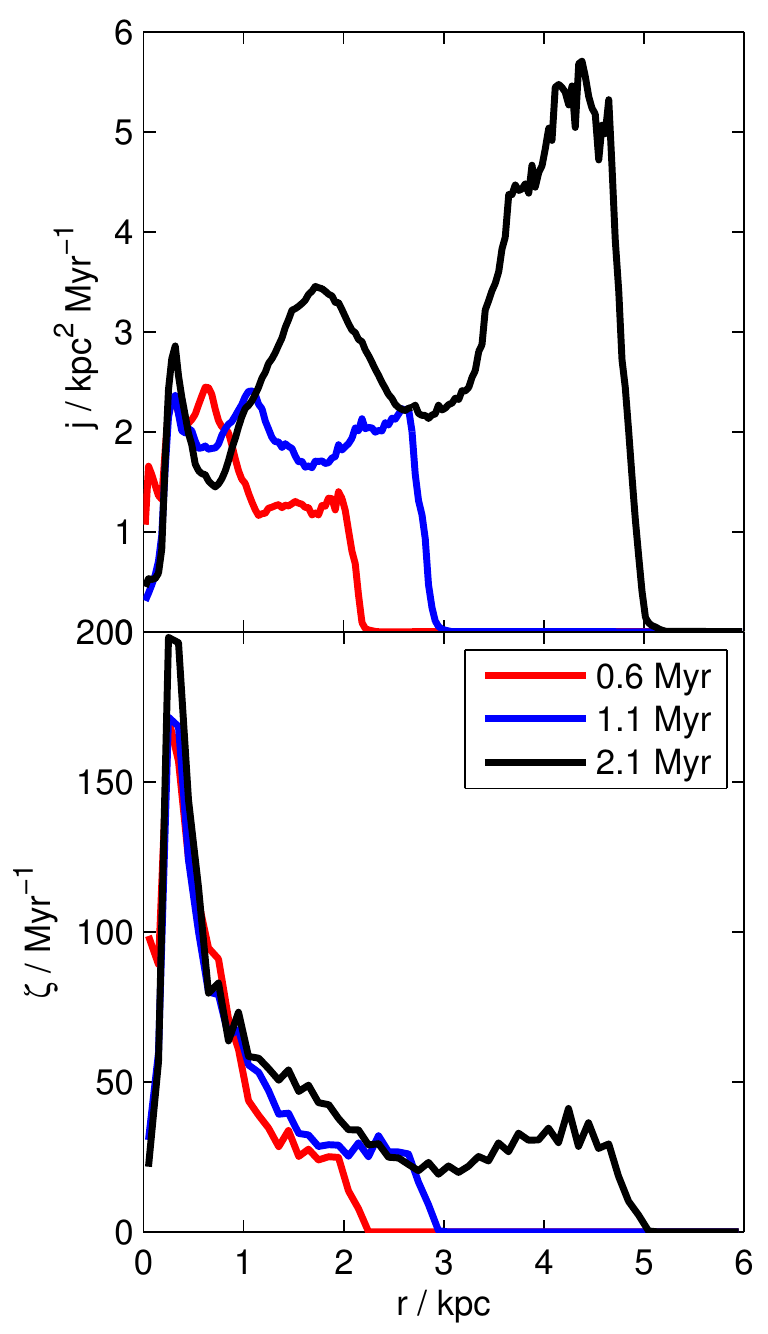}
    \caption{Magnitudes of specific angular momentum $\mathbf{j}=\mathbf{r}\wedge\mathbf{v}$ and vorticity $\mathbb{\zeta} = \mathbb{\nabla} \wedge \mathbf{v}$, in spherical bins for run dj-250p6, at $t_{age}=0.6$, $1.1$ and $2.1$ Myr; the snapshots are the same as Figure \ref{fig:velField}.}
    \label{fig:jZeta}
\end{figure}

In all the runs, we observe backflows reaching the central kpc for no longer than $\simeq2$ Myr, while the model predicts  significant activity throughout the whole life of the jet/cocoon system. It is important to notice that the model was designed from the results of 2D simulations. The reason for this discrepancy is then to be found in the changes occurring when adding the third dimension. We propose three possible causes to account for backflow damping in 3D.

\begin{enumerate}
	\item \emph{Expansion dynamics} -  In the model, the backflow originates from near the hotspot HS, from the jet gas which crosses the shock in the downstream region, due to the vorticity creation associated with the jump in the specific enthalpy (\emph{Crocco`s theorem}).\\ The efficiency of this backflow in driving gas to the central region depends however on the expansion of the cocoon: if the velocity of the hotspot is larger than the average velocity of the gas flowing back, the latter will fade away. In addition to this, the expansion of the cocoon itself results in a decreased cocoon average density. The backflow/expansion interplay is sketched in Figure \ref{fig:mass-flow_model}. In 3D, the volume expansion is faster, occurring in one more dimension. This is because those simulations were run in $d/dz$ symmetry, in order to include a non-axisymmetric cold gas component. It will not be case in axisymmetry (2.5D simulations); for instance as in \cite{Walg:2013ifa}, who study the cocoon morphology, too, though with a different equation of state for the gas.\\
    \item \emph{Large-scale vorticity} - In 3D, the gas has one more degree of freedom in its flows. So, the aforementioned vorticity on scales of the whole cocoon  may result in more gas moving in the azimuthal direction rather than flowing back to the central plane, which is the only possible flow in 2D.\\
	\item \emph{Small-scale vorticity (Kelvin-Helmholtz instability)} - Coherent backflows give rise to shear, which eventually may lead to the rise of Kelvin-Helmholtz (KH) instabilities and thus turbulent vortexes. We investigate the stability of KH modes -and whether they can destroy the flow that originated them in our simulations- in Section \ref{sub:backflowStability}. Once again, the third dimension is necessary for a correct description.
\end{enumerate}

In order to test the first two options, in Figure \ref{fig:jZeta} we plot the magnitude of the specific angular momentum $\textbf{j}=\mathbf{r}\wedge\mathbf{v}$ and the vorticity $\mathbb{\zeta} = \mathbb{\nabla} \wedge \mathbf{v}$ in spherical bins for run dj-250p6. No cocoon selection was applied this time. The three lines in the figure correspond to $t_{age}=0.6$, $1.1$ and $2.1$ Myr, i.e. the same snapshots as Figure \ref{fig:velField}. 

Direct comparison with the latter figure is revealing: three separate peaks appear in $j$. The innermost one, always at $\sim0.3$ kpc, is dominated by the jet contribution (the jet has a cylindrical velocity profile, thus $j$ becomes non-zero immediately when offset from the center), and it appears also in $\zeta$. The second one ($0.7-1.9$ kpc) is consistent with the cocoon size, and locates the bulk of the backflowing gas. We can once again see how the backflows are relevant but can hardly reach the central region. The third peak in $j$ ($1.5-4.5$ kpc) has a more irregular and extended shape, suggesting its composite origin. Indeed, contributions from both the bow shock region and the HS are present. In the bow shock, $j$ is simply consistent with (non-spherical) expansion, while the hot spot contains gas with high azimuthal velocities. 

Similar information is provided by $\zeta$; although the vorticity decays with distance from the center, a plateau of constant vorticity appears, suggesting structure on the same scales. Note that the average vorticity profile keeps constant with time, while the two outer peaks in the angular momentum distribution tend to increase their distance and magnitude. These peaks are associated with ``rings'' of highly rotating, shearing material, and it is interesting to note that these coherently rotating structures form as a consequence of the general circulation within the cocoon.

We end this overview section with some final consideration on the fate of the backflowing gas. It is true that, after about $2$ Myr, the backflow is shut down; but we have shown that by that time a large amount of hot gas has been in the innermost $1$ kpc. The final fate of this gas will depend on the dynamics of the cocoon. As is evident from Figure~\ref{fig:velField}, the cocoon continues to expand laterally, and the gas advected in the meridional plane will follow this expansion. As long as this expansion lasts, most of this gas cannot settle into, e.g., a meridional disc. Our simulations lack sufficient spatial and temporal resolution to state whether a geometrically thin accretion disc may form around the SMBH. 

However, if even a few percents of this advected gas mass can reach the innermost $100$ pc (which is, again, likely yet hard to say from our results), it will ultimately contribute to raise the total amount of gas available for accretion onto the central BH. Thus, we argue that the gas in the meridional plane is likely to supply the accretion disc around the central BH, ultimately contributing to powering the jet itself.

\subsection{Backflow stability}  \label{sub:backflowStability}
As we have previously noticed, in all the runs we have presented the backflow tends to disappear after some time. We have listed different physical mechanisms that can act to destabilize the backflow: here however we will focus our attention on the Kelvin-Helmholtz instabilities at the interfaces between the backflow and the bow shock on one side, the backflow and the jet on the other side. These interfaces are regions of very high shear, due to the negative velocity of the backflow w.r.t. both the jet and the bow shock.

The dispersion relation for the linearised KH instability in a compressible fluid is given by \cite{RevModPhys.40.652}:
\begin{equation}
\frac{x^{2}-1}{x^{4}}=a\frac{(x-m)^{2}-b}{(x-m)^{4}}\label{eq:bfs:1}
\end{equation}
where we have  assumed a form: $\exp i(kr-\omega t)$ for the perturbations and defined: $x=\omega/c_{s}k$ and: $m=V\cos\phi/c_{s}$ ($V$ being the relative velocity between the backflow and one of the two interfaces, while $\phi$ is the angle $V$ makes with such interface, and $c_s$ is the sound speed). The coefficients $a$ and $b$ are defined as:
\[
a=\left(\frac{\Gamma_{b|J}}{\Gamma_{c}}\right)^{2}\left(\frac{c_{c}}{c_{b|j}}\right)^{2},\quad b=\left(\frac{c_{c}}{c_{b|j}}\right)^{2}
\]
and we have defined the polytropic indexes $\Gamma_{c}, \,\Gamma_{b|j}$ for the cocoon, bow shock and jet, respectively, as well as the sound speeds: $c_{c}, \, c_{b|j}$. 
 
The typical temperatures in the cocoon and in the jet exceed $T\sim10^{10}$K, thus we will adopt for these regions a \emph{relativistic} equation of state (EoS). More specifically, we adopt the fits to the multispecies relativistic EoS given by \citet{2009ApJ...694..492C}, so that the sound speed will be given by their eq. (5k):
\begin{equation}
c_{c,j}=c\left(\frac{2\Gamma\Theta}{f(\Theta|\xi)+2\Theta}\right)^{1/2}\label{eq:bfs:2}
\end{equation}
where:$\Theta=kT/m_{e}c^{2}\simeq1.686\, T_{10}$, $T_{10}=T/10^{10}$K, $f=e/(n_{e^{-}}m_{e^{-}}c^{2})$ is the scaled internal energy density,
and $\xi=n_{p^{+}}/n_{e^{-}}$ is the proton/electron density ratio. \citeauthor{2009ApJ...694..492C} propose an approximation for $f$ (their
eq. 5g), namely:
\begin{equation}
f=(2-\xi)\left[1+\Theta\frac{9\Theta+3}{3\Theta+2}\right]+\xi\left[\frac{1}{\eta}+\Theta\frac{9\Theta+3/\eta}{3\Theta+2/\eta}\right]\label{eq:bfs:3}
\end{equation}
Here $\eta=m_{e^{-}}/m_{p}\sim5.44\times10^{-4}$. 

We look for an unstable mode by requiring that $x$ in eq.~\ref{eq:bfs:1} be purely imaginary: $x=iw$. Thus the left-hand side becomes real: $(w^{2}+1)/w^{4}$. The right-hand side is instead a complex expression, thus, requiring that its imaginary part be zero we arrive at the following equation:
\begin{equation}
mw\left[(m^{2}+w^{2})^{2}+(m^{2}-w^{2})(1-3b)+2b^{2}\right]=0\label{eq:bfs:4}
\end{equation}
We are not interested into the trivial neutrally stable solution $w=0$,thus we turn our attention to the term in square parentheses. By defining the reduced variable $q=w^{2}$ we finally obtain a reduced dispersion relation:
\begin{equation}
q^{2}+\left[3b-1+2m^{2}\right]q+\left[2b^{2}-\left(3b-1\right)m^{2}\right]\geq0\label{eq:bfs:5}
\end{equation}
The discriminant of this equation must be positive to obtain real solutions:
\begin{equation}
\Delta=\left(3b-1\right)^{2}+8m^{2}\left(3b-1\right)+\left(4m^{4}-8b^{2}\right)\geq0\label{eq:bfs:6}
\end{equation}
and in order to have at least one positive root one of the two following inequalities has to be satisfied:
\begin{equation}
\Lambda_{1}=3b-1+2m^{2}\leq0,\qquad\Lambda_{2}=2b^{2}-\left(3b-1\right)m^{2}\leq0\label{eq:bfs:7}
\end{equation}
We will now consider separately the two interfaces: cocoon-bow shock and cocoon-jet.
\begin{figure}
	\centering
	\includegraphics[width=1\columnwidth]{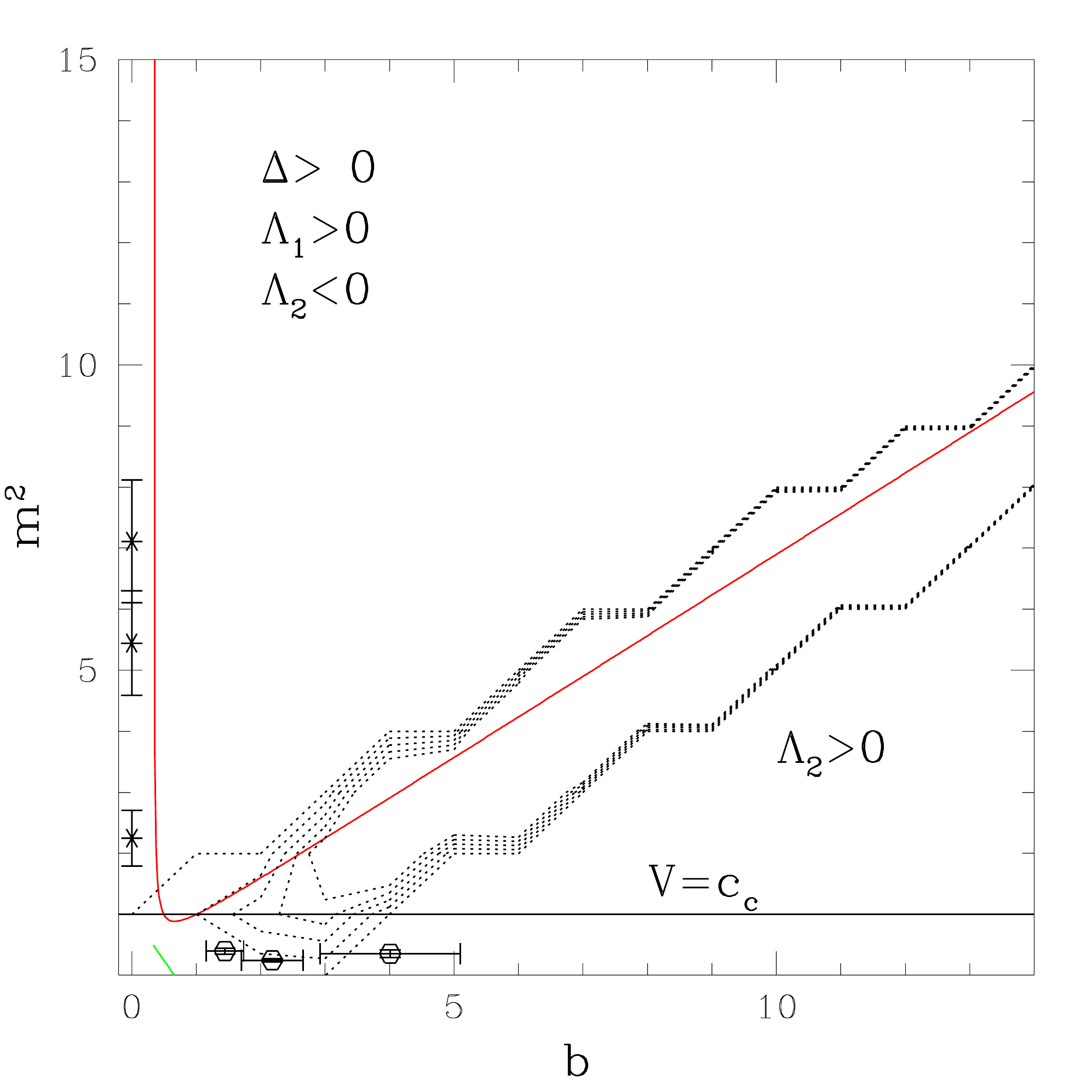}
	\caption{Conditions for existence of unstable K-H modes. The dotted contours represent values of the discriminant $\Delta=10,\, 16.62,\,23.23,\,29.85,\,36.46$ (from the outermost contour inward). The continuous curve discriminates the regions in the plane $(b,m{}^{2})$where the coefficient $\Lambda_{2}$ of eq.\ref{eq:bfs:5} changes sign. In the region where the discriminant $\Delta$ is positive and $\Lambda_{1}>0, \,\Lambda_{2}<0$ there exists real solutions of the dispersion relation. The points represent the behavior of run dj-250p6, and correspond to the same snapshots as Figure \ref{fig:velField}. Error bars represent intrinsic scatter in the gas streams. Both the \emph{straight} (starred points) and the \emph{bent} (open glyphs) backflow fall in the stable area. }
	\label{fig:bsf:1}
\end{figure}
The bow shock has relatively higher densities w.r.t. both the cocoon and the jet, and $T\aplt10^{8}$ K, thus $\Theta\ll1$ and we
can adopt a classical EoS: $p=\mu nk_{B}T$, with $\mu\sim0.62$ being the mean molecular weight of a fully ionized plasma of solar composition.
In the bow shock we will then have: $\Gamma_{b}\sim5/3$, $c_{b}=\left(k_{B}T\right)^{1/2}$.

Temperatures within the cocoon are instead larger than in the bow shock ($T_{c}\gtrsim10^{9}-10^{10}$K), thus the coefficient  $b=(c_{c}/c_{b})^{2}\gg1$, and we see that the discriminant from eq.\ref{eq:bfs:5} is positive: $\Delta\sim b^{2}\geq 0$. In the same limit we also have: $\Lambda_{1}\approx 3b+2m^{2}\geq 0$ and $\Lambda_{2}\leq 0$ for $m^{2}\gtrsim 2b/3$, i.e.: $V \gtrsim c_{c}^{2}/c_{b} $. Thus, the interface between the bow shock and the backflow can be unstable only for very large backflow velocities, largely exceeding the sound speed within the cocoon. One more empirical argument in this direction is that any shear between backflow and bow shock region is severely damped by the cocoon expansion, simply because it takes gas far away (i.e. $\phi=90^\circ$).

On the other hand, at the interface between cocoon and jet, for the coefficient, we have $T_{j}\sim T_{c}$, thus we have to inspect in detail the region of integrability of the above inequalities (eqs.~\ref{eq:bfs:6} and~\ref{eq:bfs:7}). 

In Fig.~\ref{fig:bsf:1} we show the regions where inequalities~\ref{eq:bfs:6} and~\ref{eq:bfs:7} are satisfied, together with the behavior of run dj-250p6, the same asin  Figure \ref{fig:velField}, for the same $t_{age}$ shown therein, namely $0.6$, $1.1$ and $2.1$ Myr. The error bars represent intrinsic scatter inside backflow streams (this is visible in Figure \ref{fig:velField}, too).

We see that there exists a wide region of the plane $(b,m^{2})$ where $\Delta\geqq0$ and $\Lambda_{2}\leqq0$. Here, there is a real positive solution $q$ for eq.\ref{eq:bfs:5}, and consequently a real value of $w=\pm q^{1/2}$ are allowed. So being in this region would imply the existence of at least one unstable mode, where the frequency is purely imaginary: $\omega=-iq(c_{s}k)$.

As expected, the bent backflow (open glyphs in Fig.~\ref{fig:bsf:1}) appears largely KH stable, for the reasons stated above (low values of $m^2$).
It is more interesting to notice that the straight backflow is stable too; this means that the jet material is so much hotter and less dense than the cocoon, that this difference prevails for the velocity gradient (small $b$).

We conclude that the reason for the backflow stopping is not to be found in KH instabilities. The \emph{push} this gas is given in the HS is sufficient for it to travel backwards for a few tens of kpc, then eventually it gets lost in the 3D large-scale vorticity or fades while expanding with the cocoon (as we pointed out in Section \ref{sub:backflowOverview}).

\section{Discussion} \label{sec:discussion}
We have performed a series of 3D hydrodynamical simulations of bipolar jets from AGNs, covering the first few Myr of the evolution of this object. Our main aim was to study the transition stage from a compact central source (such as the \emph{Compact Steep Spectrum} sources, \emph{CSSs}, or the \emph{Gigahertz Peaked Spectrum} sources, \emph{GPSs}) to an extended \emph{Fanaroff-Riley type two} (FRII) galaxy. We have studied the geometrical and thermodynamical properties of the extended \emph{cocoon} produced during the propagation of a relativistic jet within the ISM of its host galaxy. Inspired by recent work (\citealp{2013arXiv1305.5840S}), we have explored the consequences of varying the internal jet pressure $p_{jet}$, together with its density contrast $\rho^C_{halo}/\rho_{jet}$ and the central velocity dispersion $\sigma_V$, which in turn sets the (average) halo mass and virial radius, the SMBH mass and the jet power $P_{jet}$.

Our main results may be summarized in the following points.
\begin{enumerate}
	\item We were able to follow the evolution of compact AGNs into extended sources, distinguishing three main evolutionary stages (see Figure \ref{fig:evolution}): cocoon formation (phase C), forward propagation (phase F), and lobe expansion (phase L), when the jet eventually breaks free from the cocoon that confines it. We have highlighted the connections of the expansion history to the system's internal dynamics, especially to the jet piercing its own cocoon after a few Myr. When  this happens (or not long after), the jets develop extended lobes, thus turning from a compact to an extended source, while the cocoon (now damaged) is still confined to the innermost $5$-$10$ kpc.\\
	Indeed, any supersonic jet run for long enough will eventually break through its cocoon, since the latter is expanding at the speed of sound; it just has to recover the delay accumulated in phase C (when the cocoon was already expanding but the jet was not advancing). It is however possible that the central engine is switched off before this happens; but this scenario will be different -for instance- from FRI sources, because not only the HS, but also the jets will fade and start mixing with the surrounding gas.\\
	\item We have studied the thermodynamic state and the energy balance of the jet/cocoon system in a realistic hot ISM environment. The expansion is always to a large extent an isothermal process, with the mean cocoon temperature $T_C$ rapidly converging to $1$-$2\times10^{10}$ K. The turbulent pressure $p_t$ always converges to about $10$-$20\%$ of the cocoon total pressure, thus being dynamically significant in the long term. This happens regardless of the expansion history (mainly, independently of whether self-similarity is reached or not), thus pointing to some \emph{self-regulation mechanism} dependent more on the cocoon's internal dynamics than on the geometry of the expansion. \\The energy deposition in the ISM (in the form of $pdV$ work and $TdS$ exchanged heat) always, after phase C, remains in the interval $3$-$5\%$ of the input mechanical power $P_{jet}$. Such values of this \emph{energy coupling constant} are believed to be very significant in the galaxy formation context (e.g. \citealp{2007MNRAS.380..877S,2013MNRAS.432.3381M,2012MNRAS.424..190G}).\\
    \item Even though our simulation setup does not allow for testing different physical compositions of the jets, the results are sensitive to the ISM/jet density contrast. In particular, we have analyzed the cocoon geometry and expansion. We have found the cocoon's shape to be more elongated, in order of importance, for higher density contrast, higher $p_{jet}$ and higher $\sigma_V$ (Figures \ref{fig:visualCocoon} and \ref{fig:geoCocoon}). Very light jets ($\rho^C_{halo}/\rho_{jet}=1000$) show overall more regular shape and slower expansion. Their shape evolution during the expansion is more likely to reach a self-similar phase (constant axis ratio), so in this sense not very elongated cocoons which  undergo self-similar expansion favor lighter jets. A low injection pressure $p_{jet}$ may instead result in very little AGN/ISM coupling, giving rise in a few Myr to very large lobes.\\
    \item We have shown the presence of significant \emph{backflows}, i.e. gas circulation within the cocoons that is able to drive hot gas to the central kiloparsec. Such backflows are the product of the interaction between the jet and the local host galaxy's environment, and their contribution to the \emph{Advection Dominated Accretion Flow} (ADAF) on to the central BH demonstrates that a connection between \emph{galaxy-scale feedback} and \emph{central accretion} develops over time-scales of the order of $\sim10^{5}$ years, $\aplt1/10$ of the AGN duty cycle. This \emph{backflow accretion }time-scale is much smaller than that suggested by 2D models (\citealp{Antonuccio-Delogu:2007oq}), due to the different cocoon expansion rates and behaviour of \emph{large-scale vorticity} in 3D. We investigated the possible rise of Kelvin-Helmholtz instabilities, but found that backflows  have insufficent shear to be unstable. \\ Though the accretion timescale we find may seem small, it only refers to the typical time for the backflow to feed the ADAF, which by that time may have accreted, as we find, up to $2\times10^5\,\rm{M_\odot}$ of gas. This phenomenon thus points to a deep connection between AGN feedback and SMBH accretion, as previously hinted by \cite{2008NewAR..51..733N}. Finally, we notice that observational evidence for backflows has also been recently found by \citet{2012MNRAS.424.1149L} for two Fanaroff-Riley type I sources.
\end{enumerate}

\section*{Acknowledgments}
The authors thank Nikos Fanidakis and Christian Fendt for useful discussions during the completion
of this work. Simulations have been performed on the {\sc theo} cluster 
of the Max-Planck-Institut f\"ur Astronomie at the Rechenzentrum in Garching (RZG).
Simulation visualization has also been possible thanks to RZG facilities.
The software used in this work was in part developed by the DOE NNSA-ASC OASCR Flash Center at the University of Chicago.
AVM and AVD acknowledge travel support from the Sonderforschungsbereich 
SFB 881 `The Milky Way System' (subproject A1) of the German Research Foundation (DFG).
The work of V.A.-D. was supported by the COST Action MP0905 (`Black Holes in a Violent Universe') and by the HPC-Europa2 project (number: 1155) with the support of the European Commission - Capacities Area - Research Infrastructures.

\bibliographystyle{mn2e}
\addcontentsline{toc}{chapter}{\bibname}\bibliography{myBibliography}

\appendix

\section{Multimedia}
Movies and pictures of the simulations presented in this work are available on the S.C.'s webpage:
http://www.mpia-hd.mpg.de/$\sim$cielo/(http://mnras.oxfordjournals.
org/lookup/suppl/doi:10.1093/mnras/stu161/-/DC1).
Please note: Oxford University Press are not responsible for the content or functionality of any supporting materials supplied by the authors. Any queries (other than missing material) should be directed to the corresponding author for the article.
\label{lastpage} 
\end{document}